\documentclass[structabstract]{aa}
\usepackage{graphicx}
\usepackage[hyperindex]{hyperref}
\usepackage{amssymb}
\usepackage{aalongtable}
\usepackage{lscape}
\usepackage{natbib}
\newcommand{\gr}{{$\gamma$-ray}}
\newcommand{\lsim}{{\lower.5ex\hbox{$\; \buildrel < \over \sim \;$}}}
\newcommand{\gsim}{{\lower.5ex\hbox{$\; \buildrel > \over \sim \;$}}}

\begin{document}

\title{A complete sample of LSP blazars fully described in $\gamma$-rays}
\subtitle{New \gr\ detections and associations  with Fermi-LAT}

   \author{
          B. Arsioli  \inst{1,2,3}                 
          \and
          G. Polenta  \inst{4}
          }
        
        \institute{
Science Data Center della Agencia Spaziale Italiana, SSDC - ASI, Rome, Italy
\and
Instituto de F\'isica Gleb Wataghin, UNICAMP, R. S\'ergio Buarque de Holanda 777, 13083-859 Campinas, Brazil 
\and 
ICRANet-Rio, CBPF, Rua Dr. Xavier Sigaud 150, 22290-180 Rio de Janeiro, Brazil
\and 
ASI - Agenzia Spaziale Italiana, via del Politecnico snc, 00133 Roma, Italy \\
\email{bruno.arsioli@ssdc.asi.it,arsioli@ifgw.unicamp.br} \email{gianluca.polenta@asi.it}
                  } 


\abstract
{We study the \gr\, and broadband spectral energy distribution (SED) properties of a complete sample of 104 bright, radio-selected low-synchrotron peaked (LSP) blazars, which have well-characterized SEDs from radio to X-rays. Most of the sources have already been detected in the \gr\ band by Fermi-LAT, however almost 20\% of these blazars have no counterpart in any of the Fermi catalogs published so far.
}  
{Using the Fermi Science Tools, we look for \gr\ emission for those objects not yet reported in any Fermi-LAT catalog, finding new detections and associations. We then study the multifrequency SED for all sources in our sample, fitting their synchrotron (Syn) and inverse Compton (IC) components. A complete sample of LSP blazars with a full description in $\gamma$-ray is unique. We use this sample to derive the distribution of the Compton dominance (CD) along with population properties such as Syn and IC peak power, and frequency distributions.   
} 
{We performed a binned likelihood analysis in the 0.3-500\,GeV energy band with Fermi-LAT Pass 8 data, integrating over 7.5 years of observations. We studied \gr\ light curves and test statistic (TS) maps to validate new detections and associations, thereby building a better picture of the high-energy activity in radio-selected LSP blazars. We fit the IC component for the new detections using all data at our disposal from X-rays to GeV $\gamma$-rays, enhancing the amount of information available to study the Syn to IC peak-power correlations.  
} 
{
We deliver a unique characterization in $\gamma$-rays for a complete sample of LSP blazars. We show that three previously unidentified 3FGL sources can be associated with blazars when using improved \gr\ positions obtained from TS maps. Six previously unreported \gr\ sources are detected at TS $>20$ level, while another three show TS values between 10-20. We evaluate two cases in which source confusion is likely present. In four cases there is no significant \gr\ signature when integrating over 7.5 years. Short-lived flares at $\sim$1 month scale, however, have been detected in these sources. Finally, we measure the log(CD) for the sample, which has a Gaussian-like distribution with median log(CD)$\approx$0.1, implying that on average the peak-power for the Syn and IC components in LSP blazars are similar. 
} 
{}

 \keywords{Gamma rays: Galaxies -- Galaxies: Active -- Radiation mechanism: Non-thermal}
 
 \maketitle
 
%
%

\section{Introduction}

Blazars are a particular class of jetted active galactic nuclei (AGN), corresponding to the very few cases where jets are pointing close to our line of sight \citep{PadovaniAGNwhatinname}. These objects are known for having a rather unique spectral energy distribution (SED) often characterized by the presence of two nonthermal bumps in the log($\nu$f$_{\nu} $) versus log($\nu$) plane, which extends along the whole electromagnetic window  from radio up to TeV $\gamma$-rays. Also known for their rapid and large amplitude spectral variability, usually the observed radiation shows extreme properties owing to the relativistic nature of the jets, which result in amplification effects. Blazars are relatively rare; there are $\sim$ 4000 optically identified objects in the latest blazar surveys, 5BZcat \citep{5BZcat} and 2WHSP \citep{2WHSP}, and they have been extensively studied by means of a multifrequency approach.

According to the standard picture \citep[e.g.,][]{giommisimplified}, the first peak in log($\nu$f$_{\nu}$) vs. log($\nu$) plane is associated with the emission of Syn radiation owing to relativistic electrons moving through the magnetic field of the collimated jet. The second peak is usually understood as a result of inverse Compton (IC) scattering of low energy photons to the highest energies, by the same relativistic electron population that generates these Syn photons (synchrotron self Compton model; SSC). The seed photons undergoing IC scattering can also come from outside regions, such as the accretion disk and broad line region, and can add an extra ingredient (external Compton models; EC) for modeling the observed SED.  

Since the peak-power associated with the Syn bump tell us at which frequency ($\rm \nu_{peak}^{Syn}$) most of the AGN electromagnetic power is being released, the parameter log($\nu_{peak}^{Syn}$) has been extensively used to classify blazars. Following \cite{padgio95,abdo10}, objects with log($\nu_{peak}^{Syn}$) $<$ 14.5, in between 14.5 to 15.0, and $>$ 15.0 [Hz] are called low, intermediate, and high Syn peak blazars, i.e., LSP, ISP, HSP, respectively.

Blazars are by far the largest population of high galactic latitude sources in all Fermi-LAT catalogs and all types of blazars (LSP, ISP, and HSP) have been detected in $\gamma$-rays. Nevertheless a relatively large percentage of LSPs still lack detection by Fermi-LAT. The fact that some of these LSPs are relatively bright in radio and show hints of distinct optical polarization properties \citep{GammaQuietBlazars2016MNRAS} has motivated arguments about the existence of a specific class to represent the so-called \gr\ quiet blazars; the latter have a relatively lower polarization fraction. Therefore, current evidence gives us a hint of the jet condition regarding $\gamma$-ray undetected blazars, showing they might be connected to relatively less magnetized jets. In particular, \cite{Blinov-2015} probed optical polarization swings in connection to periods of enhanced activity in $\gamma$-rays, therefore in connection with magnetic field strength and ordering within the jet structure. But the  observations of orphan $\gamma$-ray flares, which have no counterparts at the low-energy site (optical to X-ray), are a standing challenge for current SSC and EC models when trying to describe blazar variability \citep{Potter-2018-Orphan-Flares}. Alternative models such as the Ring of Fire \citep{McDonald-2017-Sheath-Orphan-Flares} discuss evidence for a sheath of plasma surrounding the spine of the jet, producing a dominant IR photon field that would undergo IC scattering.  Despite that, other works recognize $\gamma$-ray undetected blazars as MeV peaked \citep{MeVPeaked2017}, which could be out of reach for Fermi-LAT.  

As known, all Fermi-LAT official catalogs are blind with respect to other wavebands, meaning that information about sources detected in other energy bands are not taken into account. Each new \gr\ detection must meet the conservative requirement of being observed at least as a 5\,$\sigma$ excess compared to the expected background. This is usually incorporated in the so-called test statistic (TS) parameter (more on sec. \ref{TS}) translating to a TS\,$>$\,25 requirement for acceptance of new \gr\ sources. This was necessary to avoid spurious detections and misleading associations, especially pre-Fermi-LAT, when the main population of \gr\ emitters was still to be identified.

The situation now has changed since the astrophysics community already recognize blazars as the main population of extragalactic \gr\ emitters. Searches for \gr\, emission in samples of previously known blazars can relax the TS\,$>$\,25 requirement, as discussed in \cite{1BIGB}. These authors tested the existence of \gr\, signatures by including more information compared to a pure blind approach and they indeed detected 150 new \gr\ sources. The detection of new sources, together with resolving cases of source confusion, should not be neglected given its direct impact on statistical \gr\ properties of specific populations, particularly for small samples.

Especially in case of LSP blazars, the percentage of $\gamma$-ray undetected sources is higher. On average, those blazars have steep \gr\ photon spectral index \citep{3LAC} that can compromise the detection of high redshift sources since the \gr\ SED shifts to lower frequencies in the observer rest frame. Also, the absorption of very high energy (VHE) \gr\ owing to the interaction with the extragalactic background light \cite[EBL;][]{EBLabsorption} may hinder observations with Fermi-LAT. In addition to those possible complications, intrinsic jet properties (such as the Doppler and beaming factor  $\delta$) and the dominant IC regime (either SSC or EC) may also have a large influence on the Fermi-LAT detectability of radio-loud blazars. In fact, \cite{RadioLoudDoppler-Beamed} showed that the \gr\ sources detected during the first three months of Fermi-LAT operations are associated with the largest apparent jet speeds (therefore the largest bulk Lorentz factors) as deduced from radio measurements with the Very Large Baseline Array (VLBA). Also, \cite{Lister2015} showed that the $\gamma$-ray detection of LSP blazars is relatively less likely when the Lorentz factor is low and the Syn peak is below 10$^{13.4}$Hz. In fact, Sec.~\ref{flaring} presents three examples of radio-loud blazars with relatively bright $\nu$f$_{\nu}$ Syn peaks that are detectable only during short flare episodes. Those cases could provide hints of the mechanisms behind \gr\ flares and are interesting targets for upcoming missions such as the MeV dedicated e-Astrogram \citep{eASTROGAM2016proceedings,eASTROGAM2017}.  

Motivated by the possibility of unveiling new \gr\ sources, we use a complete sample of radio-loud blazars, and consider all cases not yet detected at GeV band for a likelihood analysis with the Fermi Science Tools. Our approach shows that most of the previously \gr\ undetected sources are actually detectable when integrating over 7.5 years of observations or during short periods of their flaring states. We present \gr\ light curves adopting a one month time bin, showing that currently undetected blazars may introduce a dynamic high-energy component to the sky with numerous short periods of $\gamma$-ray activity. This might build a considerable portion of the extragalactic \gr\ background \citep[EGB; ][]{dgrbFermi,Fermi-Background-Var-2013} whose origin is still being debated \citep{NatureIDGB,EGRB-Composition,DifuseGammaBLlacs,50GeVbackground}. In addition, undetected blazars may add anisotropic contributions to the EGB content especially at small angular scales\citep{GammaAnisotropyFermiTeam,GammaAnisotropy,blazarbackground} thus potentially affecting the search for dark matter annihilation or decay signatures in connection to diffuse \gr\ emission from extragalactic large-scale structures \citep{dgrb-LSS,CountGammaCluster}. Indeed, our new detections may complement current understanding of the EGB origin especially for the MeV band where LSP and ISP blazars are more relevant.

Our detailed search for new \gr\ sources also provides a better description of the IC component for many blazars that so far had limited or no \gr\ data available, thus resulting in a measurement of Syn and IC peak-power parameters for nearly all sources (except for five nondetections). Since this radio sample includes the brightest and well-characterized LSP blazars, we study population properties such as the distribution of Syn and IC peak frequencies and peak power, and the distribution of log($\nu_{IC}/\nu_{Syn}$) and log(Compton dominance).

\begin{table*}
\centering
\caption{List of new $\gamma$-ray detections and associations for the Radio-Planck sources, following a case by case description in  Sec.~\ref{new}. Columns Source and Redshift list the blazar name and distance as from \cite{5BZcat}. We note that Table \ref{tableRadioPlanck} brings an extra name reference as from NVSS Radio Catalog \citep{NVSS} given that each blazar has a well-defined radio counterpart. Columns $ \rm log(\nu^{Syn}_{peak})$ and log($\nu^{Syn}_{peak}$) refer to the peak frequency from Syn and IC components measured in Hz. Columns log($\nu f^{Syn}_{\nu}$) and log($\nu f^{IC}_{\nu}$) correspond to the Syn and IC peak-power measured in erg/cm$^2$/s. Column N$_0$ list the pre-factor as of eq.~\ref{powerlaw} in units of ph/cm$^{2}$/s/MeV when assuming E$_0$ = 1000\,MeV. Column $\Gamma$ list the photon spectral index as of eq.~\ref{powerlaw}. Column TS list the test statistic values associated with the $\gamma$-ray signature for each source when integrating over 7.5 years of Fermi-LAT PASS8 observations. All model parameters were derived using the Fermi Science Tools assuming a power-law spectrum within the 0.3-500\,GeV energy band. New detections at $\rm TS>20$ level are denoted with an ``a" flag; new association from better positioning of 3FGL sources are given a ``b" flag, faint detections with TS between 10-20 are given a ``c" flag; sources only detectable during short flare episodes are given a ``d" flag (for those cases, TS values refer to the bright state integrated over one month); and sources that could be confused with a second \gr\ emitter are given an ``e" flag.}  
\begin{tabular}{cccccc|ccc}
Source & Redshift & log($\nu_{peak}^{Syn}$) & log($\nu$f$_{\nu}^{Syn}$) & log($\nu_{peak}^{IC}$) & log($\nu$f$_{\nu}^{IC}$)& $N_0$ (10$^{-13}$) & $\Gamma$   & TS$^{Pass \ 8}_{7.5 \ yrs}$   
\\
\hline

5BZQJ\,0359+5057$^{a}$  & 1.512 & 12.1      & -10.7 & 21.3 & -10.4 & 36.6$\pm$1.5  & 2.72$\pm$0.04   & 786.0  \\

5BZQJ\,1955+5131$^{a}$  & 1.210 & 12.6      & -11.7 &  20.8 & -11.1 & 7.0$\pm$0.7  & 2.56$\pm$0.10  & 119.7  \\

5BZUJ\,0909+4253$^{a}$  & 0.670 & 12.9      & -11.5 & 20.9 & -11.7 & 3.5$\pm$0.5  & 2.62$\pm$0.15   & 74.4  \\

5BZQJ\,1153+8058$^{a}$  & 1.250 & 12.6      & -12.0 & 21.1 & -11.9 & 2.1$\pm$0.4  & 2.53$\pm$0.19   & 33.7  \\

5BZQJ\,0646+4451$^{a}$  & 3.396 & 11.6      & -11.8 & 20.8$^?$ & -11.7 & 3.8$\pm$0.7  & 3.39$\pm$0.24    & 85.6  \\

5BZQJ\,0125-0005$^{a}$  & 1.077 & 12.8      & -11.6 & 20.3 & -11.6 & 2.2$\pm$0.5  & 2.90$\pm$0.31   & 22.5  \\ 

5BZQJ\,0555+3948$^{b}$  & 2.365 & 12.0      & -11.8 & 20.8 & -10.9 & 10.2$\pm$1.0  & 3.12$\pm$0.11   &  195.5 \\

5BZUJ\,0433+0521$^{b}$  & 0.033 & 13.8      & -10.2 & 19.8 & -10.2 & 14.4$\pm$1.0  & 2.97$\pm$0.08   &  370.4 \\

5BZQJ\,0228+6721$^{b}$    &  0.523  &  12.8   & -11.2 & 21.1 & -11.4  &   8.0$\pm$1.0 & 2.65$\pm$0.12  &  79.4  \\

5BZQJ\,2218-0335$^{c}$  & 0.901 & 12.3      & -11.3 & 21.0$^{?}$ & -11.7 & 2.7$\pm$0.6   & 2.80$\pm$0.27  & 19.9  \\

5BZBJ\,0006-0623$^{c}$  & 0.347 & 13.0      & -11.1 & 20.3 & -11.9 & 1.4$\pm$0.4   & 2.13$\pm$0.20  & 17.5  \\

 
5BZQJ\,1038+0512$^{c}$  & 0.450 & 12.0      & -11.8 & 20.8$^{?}$ & -12.1 & 1.6$\pm$0.6   & 2.79$\pm$0.38  & 10.1  \\

5BZQJ\,0010+1058$^{d}$  & 0.089 & 14.5      & -10.7 & 20.5 & -10.8 & 4.1$\pm$1.9  & 3.22$\pm$0.53  & 26.0$^{*}$  \\


5BZUJ\,0241-0815$^{d}$ & 0.005 &  13.9$^{?}$  & -10.3  & 20.9$^{?}$  & -10.6  & 6.3$\pm$0.2  & 1.64$\pm$0.01  & 12.0$^{*}$  \\

5BZQJ\,2136+0041$^{d}$ & 1.941 & 10.8 & -11.7  & 21.2  & -11.2 &  19.9$\pm$2.0 & 2.46$\pm$0.10 &  7.5$^{*}$  \\

5BZQJ\,1642+6856$^{e}$  & 0.751 & 12.5   & -11.6 & 20.1$^{?}$ &  -12.3  &  9.1$\pm$4.0  & 2.31$\pm$0.28 &  8.3   \\


\end{tabular}
\label{detectiontab}
\end{table*}

\section{Sample description}

We consider the complete sample of radio-loud AGN that was studied in detail by \citet{PlanckEarlySEDs}, consisting of 104 northern and equatorial sources with declination larger then -10$^\circ$ and  flux density at 37\,GHz exceeding 1\,Jy as measured with the Mets\"ahovi radio telescope. We refer to those sources as the Radio-Planck sample and list these sources in table \ref{tableRadioPlanck} showing their blazar names from 5BZcat \citep{5BZcat}, and their NVSS\footnote{National Radio Astronomy Observatory Very Large Array Sky Survey, the so-called NVSS Catalog} radio counterpart. Out of these sources, 83 have a confirmed \gr\ counterpart in at least one of the Fermi-LAT \citep{FermiLAT} catalogs, i.e., 1FGL, 2FGL, and 3FGL  \citep{1FGL,2FGL,3FGL} or 2FHL and 3FHL \citep{2FHL,3FHL}. However, the nondetection by Fermi-LAT of the remaining 21 equally bright radio-loud AGN with similar radio-to-optical SED is intriguing, rising discussions on both the nature of the high-energy emission in blazars and the efficiency of Fermi-LAT in solving faint \gr\ sources \citep{Lister2015}. It has been argued that LSP blazars with $\nu_{peak}<$\,10$^{13.4}$ Hz may show a typical  IC peak below 0.1\,GeV out of the Fermi-LAT sensitivity bandwidth (0.1-500\,GeV), so that we can only probe the very end of the IC component \citep{MeVPeaked2017}. 

\section{ \gr\ analysis of undetected sources } 
\label{TS}

In a blind analysis, the spectral parameters of a hypothetical source, such as normalization and photon spectral index, and also the source position itself (R.A. and Dec.) are all free parameters that have to be optimized during the data analysis. We might be able to reduce the uncertainty, however, with respect to position since we know multiple $\gamma$-ray blazar candidates that have already been identified from multifrequency observations from radio up to X-rays. Actually, a total of 21 sources in our Radio-Planck sample do not have a \gr\ counterpart in previous Fermi FGL catalogs. Those constitute a set of 21 seed fixed positions, for which we test the existence of relevant $\gamma$-ray signatures. This method has been successfully tested for a set of 400 $\gamma$-ray candidates \citep{1BIGB} preselected from a sample of HSP blazars, resulting in 150 new detections.

We performed a likelihood analysis integrating over 7.5 years of Fermi-LAT observations in the 0.3-500\,GeV band using Pass 8 data release \citep{Pass8Description}, and assuming the \gr\ spectrum of a new source could be described by a power-law model as
\begin{equation} \rm
\frac{dN}{dE}  =  N_0  \Big( \frac{E}{E_0} \Big)^{-\Gamma}
\label{powerlaw}
,\end{equation} 
where $\rm E_0$ is a scale parameter (also known as pivot energy), $\rm N_0$ is the pre-factor (normalization) corresponding to the flux density in units of ph/cm$^{2}$/s/MeV at the pivot energy $\rm E_0$, and $\rm \Gamma$ is the photon spectral index for the energy range considered. Both $\rm \Gamma$ and $\rm N_0$ are set as free parameters and further adjusted by the fitting routine \textit{gtlike}. Source positions and $\rm E_0 = 1000\,$MeV are set as fixed  parameters, therefore constants for the likelihood analysis. In the source-input xml file, all sources within 10$^\circ$ from the candidate had both $\rm \Gamma$ and $\rm N_0$ parameters flagged as free\footnote{In this regard we are following recommendations from Fermi Science Tools user guide \url{https://fermi.gsfc.nasa.gov/ssc/data/analysis/scitools/binned_likelihood_tutorial.html}, which recommends setting free parameters at least within 7$^\circ$ from the source of interest. This is a consequence of the large point spread function (PSF), especially at low-energy threshold, which can overlap with nearby sources. Therefore, in order to get a confident description of a particular source, we properly fit and adjust the whole environment around it.}. Therefore the 3FGL models of these sources, which are based on four years of observations were adjusted, since we integrated over 7.5 years of data. This particular choice increases the computational burden of the analysis, but it is crucial for adapting the model maps to the extra 3.5 years of exposure that is being considered. Results are shown in  Table~\ref{detectiontab}, listing only the 16 cases that had no known counterpart from previous $\gamma$-ray catalogs and which we now describe case by case in this work. Also we note that Table~\ref{tableRadioPlanck} holds the description for the entire Radio-Planck sample.

Since we are dealing with sources that are predominantly LSP blazars, we should expect them to have steep $\gamma$-ray SED (in log($\rm \nu f_{\nu}$) versus log($\nu$) plane) as observed from the correlation in $\rm \Gamma$ versus $\rm log(\nu^{Syn}_{peak})$ plane \citep{3FGL}, where LSP sources dominate the $\rm \Gamma > 2.0$ side. The parameter $\rm N_0$ represents the flux at 1\,GeV (given our choice for $\rm E_0$). Therefore new detections are expected to be on the border or below Fermi-LAT four year sensitivity limit\footnote{3FGL sensitivity considering the instrument response function (P7REP-SOURCE-V15), as described at Fermi-LAT repository \url{http://www.slac.stanford.edu/exp/glast/groups/canda/archive/p7rep_v15/lat_Performance.htm} }, which is $\rm E^2 dN/dE \approx 2.0 \times 10^{-12}$ erg/cm$^2$/s at 1\,GeV, i.e., $ \rm dN/dE \approx 12.5 \times 10^{-13}$ ph/cm$^2$/s/MeV. Following our results as reported in Table~\ref{detectiontab}, few cases have $\rm N_0$ above the the Fermi-LAT four year sensitivity level. In particular 5BZQJ\,0359+5057 and 5BZUJ\,0241-0815 have larger $\rm N_0$, which is clearly a consequence of an enhanced $\gamma$-ray activity reported just after 2013 (out of the integration period used to build the 3FGL catalog) as discussed in Sec. \ref{notes}  and \ref{flaring}. Another three sources flagged with ``b" have $\rm N_0$ close to the 3FGL sensitivity border; those are actually detected in the 3FGL catalog but with large position uncertainty, which lead them to be unassociated in previous catalogs.

Positive \gr\ signatures were first evaluated based on test statistics (TS) values as defined by \cite{mattox}: $ \rm TS = -2ln \Big(  \frac{L_{(no ~ source)}}{L_{(source)}} \Big)$, where $\rm L_{(no ~ source)}$ is the likelihood of observing a certain photon count for a model without the candidate source (the null hypothesis), and $\rm L_{(source)}$ is the likelihood value for a model with the additional candidate source at the given location. The reported TS values correspond to a full band fitting, which constrains the whole spectral distribution along 0.3-500\,GeV to vary smoothly with energy and assuming no spectral break. Considering that we have a good description of the Galactic and of the extragalactic diffuse components, this is a measure of how clearly a source emerges from the background, also assessing the goodness of free parameters fit.

A TS\,$\approx$\,25 is equivalent to a 4-5$\sigma$ detection, depending on the strength of the background in the region \citep[][]{1FGL}, and only cases with TS\,$>$\,25 are considered by the Fermi-LAT team as a positive detection of a point-like source. Following the discussion on \cite{1BIGB}, we analyzed \gr\ signatures down to TS=10, which are spatially consistent with blazars already known from other energy bands and double checked these with TS maps.

A TS map consists of a pixel grid where the existence of a point-like source is tested for each pixel, and each grid bin is evaluated using a likelihood analysis \footnote{For a complete description of the Fermi Science Tools, check: \url{https://fermi.gsfc.nasa.gov/ssc/data/analysis}.}. Since the PSF improves with energy, we worked with E$>$500\,MeV photons to help us determine the TS peak position with better precision than working at lower Fermi-LAT bandwidth (down to 100\,MeV). Thus the map alone tests the existence of a point-like source  emerging from a flat low-TS background. 

We enhanced the \gr\ characterization of the Radio-Planck sample by $\approx$15$\%$, since we now describe ten new detections of steady sources as discussed in Sec.~\ref{notes}, Sec.~\ref{confusion}, and Sec.~\ref{faintdetec}; three new associations from improved position of previously known 3FGL sources as discussed in Sec.~\ref{improveposition}; and three detections of sources with isolated flaring activity as discussed in Sec.~\ref{flaring}. In the following discussion we present fitting parameters, SEDs, light curves for most relevant cases, and TS maps for all sources. We comment on four cases where we find poor evidence of \gr\ signature, which is probably related to high redshift, IC peak at MeV band, and low galactic latitude hindering the detection.

For the light curves, we usually considered a time bin of approximately 30 days, and estimated the corresponding flux and errors only when the TS per bin was larger than 4.0. When this condition was not satisfied, an upper limit to the flux was calculated using the integral method (provided by the Fermi Science Tool), which takes into account the background level and spectral properties of the test source.

\section{New \gr\ detections, validation, and association}
\label{new}

We present new \gr\ signals down to TS\,$\approx$\,10 level. We also describe one case of source confusion, solve three cases of unassociated 3FGL sources, and comment on the nondetections as well. Table~\ref{detectiontab} shows the power-law parameters resulting from the fit in the 0.3-500\,GeV energy band, together with redshift of the counterpart, Syn, and IC peak frequency ($\nu_{peak}$), and flux density ($\nu$f$_{\nu}$) for all cases studied. For each source, we present a TS map together with the \gr\ SED with a polynomial fit for both Syn and IC component. Following \cite{giommisimultaneous}, when fitting the nonthermal component we also account for optical and UV thermal emission due to accretion using the composite optical spectrum built by \cite{FSRQ-template}. This thermal template is based on 2200 optical spectra of radio-quiet quasars (QSOs) taken from the SDSS database and its expected soft X-ray emission, from \cite{Grupe2010}.

The TS maps are calculated considering only photons with E\,$>$\,500\,MeV, which is a good choice to evaluate the TS spatial distribution for these radio sources, since they usually have a photon spectral index in the range 2.0\,$< \Gamma <$\,3.0. Whenever possible, we also use E\,$>$\,1.0\,GeV photons (or higher) to improve the localization of \gr\ signatures. We call attention to the fact that for most cases we are dealing with relatively faint $\gamma$-ray sources, therefore the TS distribution could peak at a position slightly offset from it counterpart. An offset is expected given the uncertainty introduced by the large PSF, which is on the order of 1.4$^\circ$ at 500\,GeV  \citep{Pass8Description} and 0.81$^\circ$ at 1\,GeV level. Following the discussion in \cite{1FGL} what is important to ensure a proper match is that the 68\% confinement region for the $\gamma$-ray signature should enclose the counterpart blazar. This is explicitly shown case by case to help validate our new detections and associations.

We show the light curves with monthly bins for those cases for which we identify a $\gamma$-ray flaring activity during the 7.5 yrs covered by Fermi-LAT. Light curves are computed with likelihood analysis, therefore the background is extracted and flux points are calculated only for time bins that have TS\,$>$\,4. For all cases we compute upper limits and errors bars using the integral method assuming a 95\% confidence level, as provided by the Fermi Science Tools. When relevant for the discussion, we also show the significance of the \gr\ signatures based on the 3FGL setup (i.e., likelihood analysis with 4.0 yrs of Pass 7 data) to test if those sources could have been identified previously.

\begin{figure}[]
   \centering
    \includegraphics[width=1.0\linewidth]{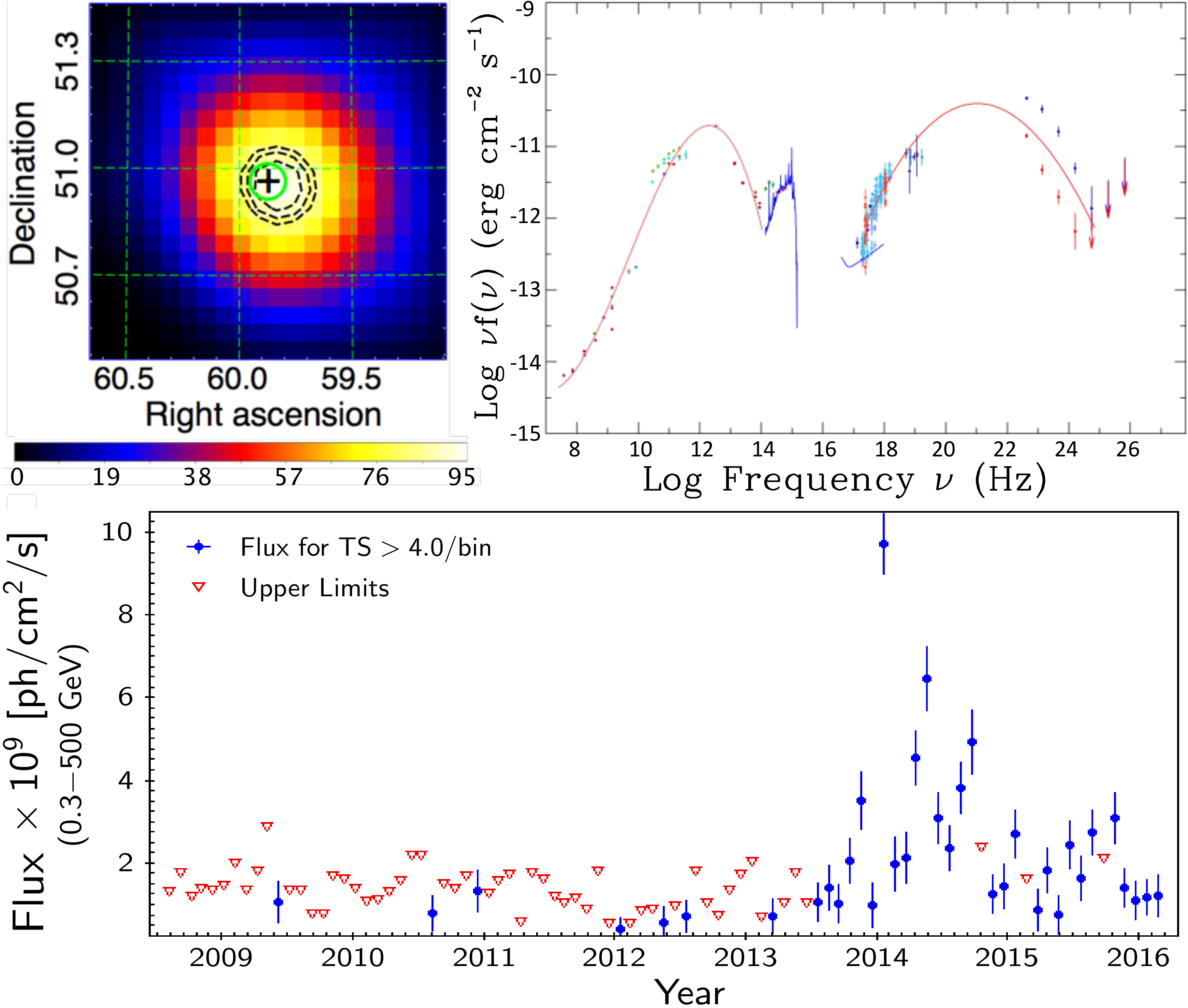}
     \caption{{\bf 5BZQJ\,0359+5057}. Top left: TS map considering only E\,$>$\,0.5\,GeV photons with black dashed lines representing 68\%, 95\%, and 99\% containment regions for the \gr\ signature (used for all further TS maps), considering the brightest period from 2014 to 2015; 5BZQJ\,0359+5057 is marked with a circle centered at ``+". Top right: The SED with a polynomial fit to the mean Syn and IC components; the blue-bump feature between 10$^{14}$ and 10$^{18}$\,Hz \citep[][]{FSRQ-template} for a source at z=1.512. In the \gr\ band, red points represent the SED before flaring (before 2013), while blue is for the flaring period (after 2013).  Bottom: 0.3-500\,GeV light curve along 7.5 yrs of observations with bins of 30 days; red points represent upper limit flux.}
      \label{bzq0359}
\end{figure}

\subsection{Notes on individual objects}
\label{notes}

We present a detailed description for the new $\gamma$-ray detection, which showed $\rm TS\,>\,20$ when integrating over 7.5 years of observations. Those cases are relatively isolated, meaning there are no other close by $\gamma$-ray counterparts that could contaminate the observed signatures. For the discussion we include TS maps, light curves, and spectral points. Together, those elements build an entire picture, not only validating and describing their $\gamma$-ray properties, but also explaining (whenever necessary) why those sources were not previously detected, showing examples on how the data treatment is refined with multifrequency information, leading to more efficient use of public databases.

{\bf 5BZQJ\,0359+5057.} Computing a TS map in the 0.5-12\,GeV energy range (Fig.~\ref{bzq0359}), we identified a bright point-like source clearly emerging from a low and flat TS background. The light curve plotted in Fig.~\ref{bzq0359}, shows that the source 5BZQJ\,0359+5057 (4C\,+50.11) was undetectable by Fermi-LAT, most of the time with the exception of the period between June 2013 and April 2016 when it underwent a phase of strong \gr\ activity. This is consistent with its nondetection up to the 3FGL catalog (since 3FGL only integrate observations from Aug. 2008 up to Aug. 2012). 

To reproduce the 3FGL setup, we performed a likelihood analysis with Pass 7 data, integrating only during the first four years of observations. As result we found TS$^{Pass \ 7}_{4.0 \ yrs}$=15.2, with model parameters N$_0$\,=\,8.1$\pm$2.2\,$\times$\,10$^{-13}$\,ph/cm$^2$/s/MeV and $\Gamma$=2.55$\pm$0.19, using E$_0$=1000\,MeV as pivot energy. Hence, this source was out of the 3FGL catalog simply because it did not meet the TS\,$>$\,25.0 criteria. We also noticed that the parameters estimated at the time of 3FGL catalog were already in good agreement with those we present on Table~\ref{detectiontab}, showing that \gr\ signatures at 10 to 25 TS level contain rich information, as discussed in \cite{1BIGB}. 

As an exercise, we used a setup similar to 3FGL (integrating over 2008-2012) but now with PASS 8 data. This results in a firm detection with $\rm TS > 25$ and the $\gamma$-ray SED shown as red points in Fig.~\ref{bzq0359} (bottom). This give us a solid idea on how the Pass 8 data release brings way relevant enhancements for the description of the \gr\ sky.

\begin{figure}[]
   \centering
    \includegraphics[width=1.0\linewidth]{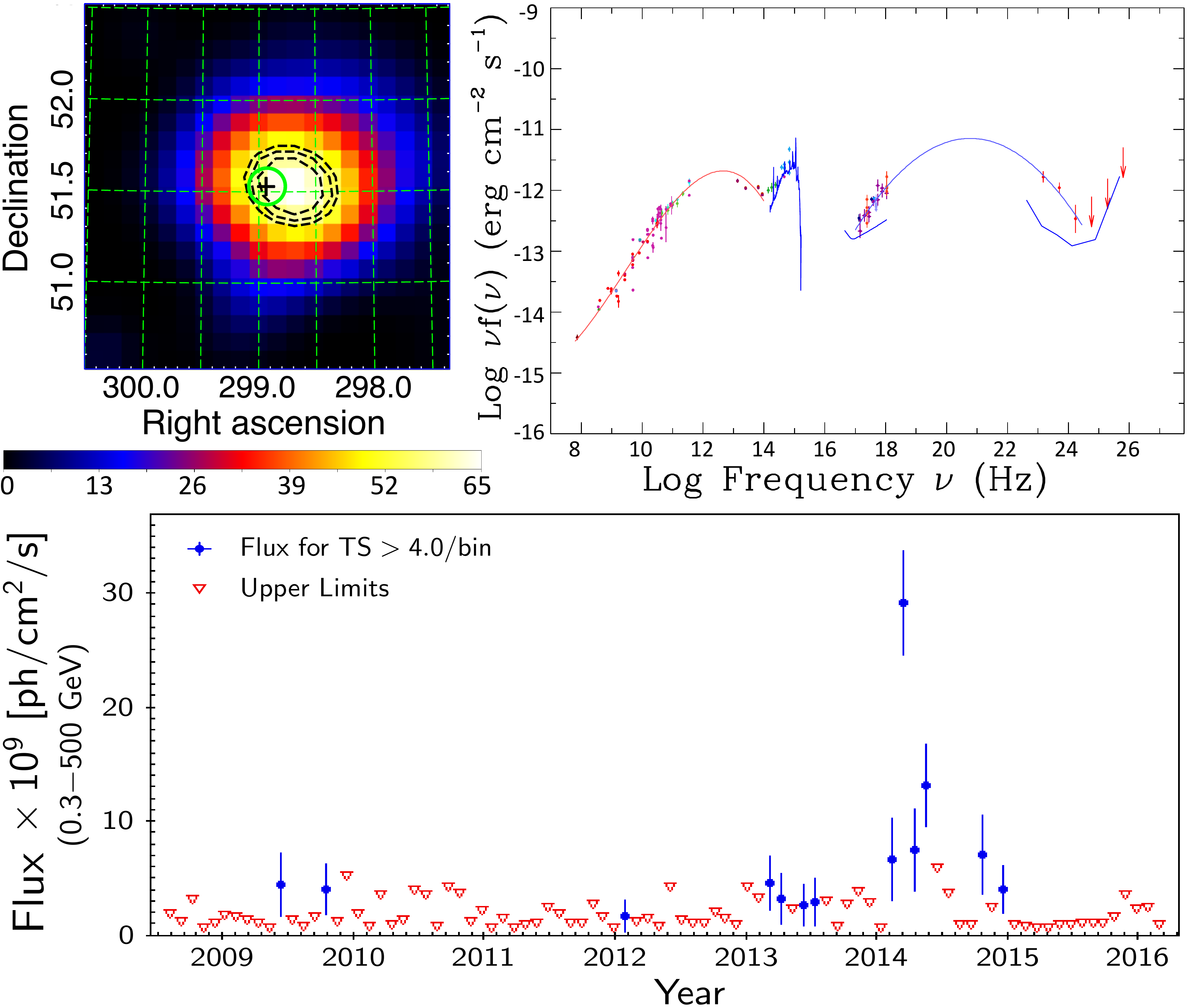}
     \caption{{\bf BZQJ\,1955+5131}. Top left: TS map considering only E$>$0.5\,GeV photons, during the high-state period between 2014 and 2015. Top right: SED for 5BZQJ\,1955+5131 at z=1.21, with the new \gr\ spectrum in the 0.3-500\,GeV band. As example, in the  10$^{22}$-10$^{26}$\,Hz band (0.1-500\,GeV) we show the Fermi-LAT sensitivity limit when integrating over four year of observations. Bottom: The \gr\ light curve for 5BZQJ\,1955+5131 along 7.5\,yrs of observations and with time bin of 30 days.}
      \label{bzq1955}
\end{figure}

{\bf 5BZQ\,J1955+5131.} This blazar shows a strong \gr\ signal when integrating over 7.5\,yrs with Pass 8 data, but from the light curve (Fig.~\ref{bzq1955}) we measured significant flaring activity only after 2014. A likelihood analysis integrating Pass 7 data from Aug. 2008 up to Aug. 2012 (hereafter we refer to this as the 3FGL setup) have shown TS$^{Pass\,7}_{4.0\,yrs} \approx$\,5.0, and therefore this source is out of the 3FGL catalog because of its variability. In addition, this blazar is close to the galactic plane (b = 11.7$^{\circ}$), where the low-energy diffuse background is more intense thus hindering detections of faint \gr\ sources. In fact, owing to larger background levels, the Fermi-LAT sensitivity in the Galactic plane region is lower than at high Galactic latitude\footnote{For a description on the Fermi-LAT performance check \url{https://www.slac.stanford.edu/exp/glast/groups/canda/lat_Performance.htm}}.

\begin{figure}[]
   \centering
    \includegraphics[width=1.0\linewidth]{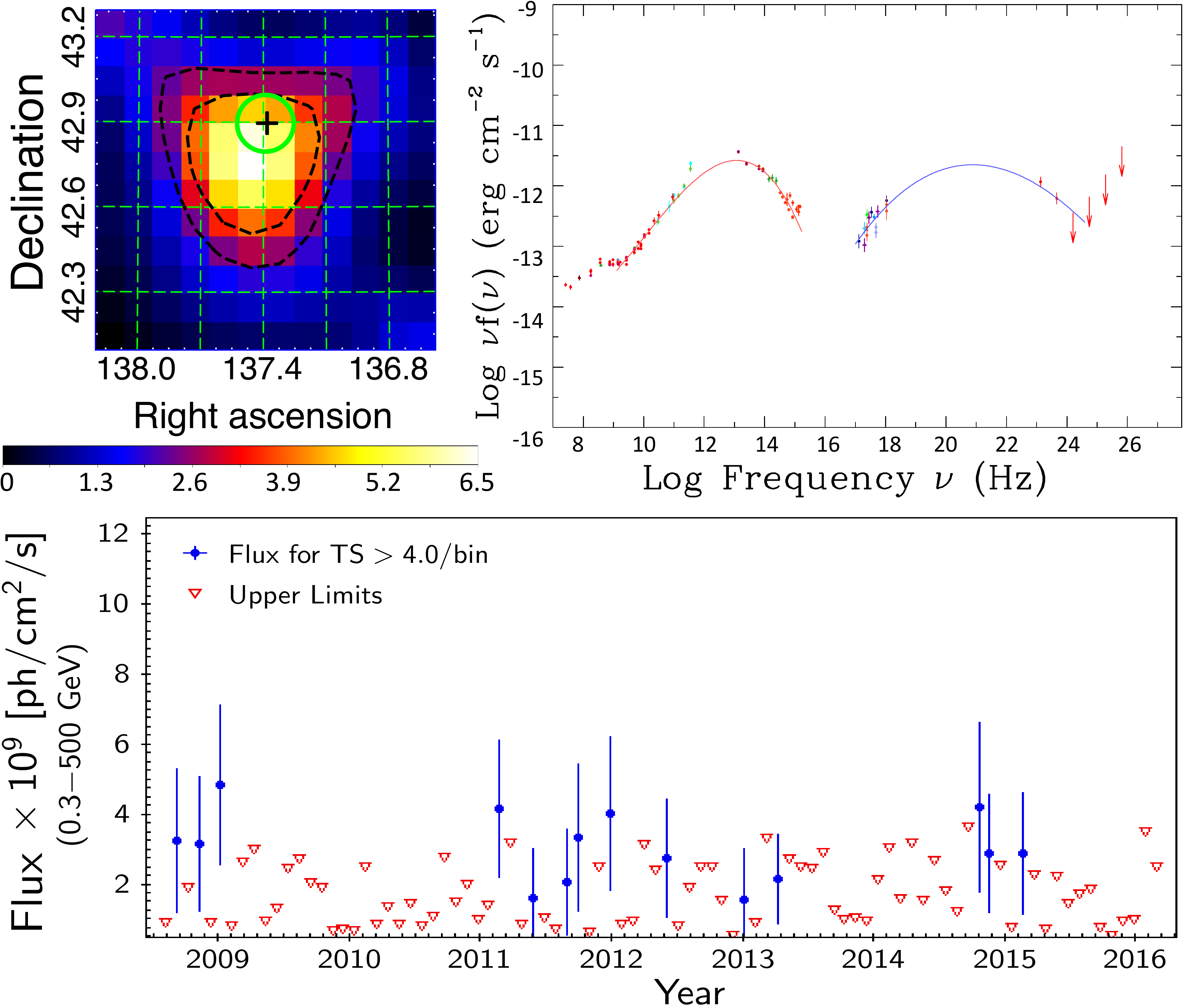}
     \caption{{\bf 5BZUJ\,0909+4253}. Top left: TS map considering only E\,$>$\,0.5\,GeV photons during the brightest period, from January 6 to February 6, 2009. Top right: SED for 5BZUJ\,0909+4253, z=0.670. Bottom: The \gr\ light curve along 7.5\,yrs of observations.}
      \label{bzu0909}
\end{figure}

{\bf 5BZUJ\,0909+4253.} This blazar is strongly detected when integrating over 7.5\,yrs with Pass 8 data (TS$^{Pass\,8}_{7.5\,yrs}$\,=\,74.4), but shows no strong flaring activity  as can be seen in Fig.~\ref{bzu0909}. A likelihood analysis with the 3FGL setup gives TS$^{Pass\,7}_{4.0\,yrs} \approx$\,14.2, with power-law parameters N$_0$\,=\,2.4$\pm$0.8\,$\times$\,10$^{-13}$\,ph/cm$^2$/s/MeV, $\Gamma$\,=\,3.04$\pm$0.38, using E$_0$\,=\,1000\,MeV as pivot energy, which agrees with values reported in Table~\ref{detectiontab}. This is another example for which it would be beneficial to have preliminary information about faint signatures with TS in between 10 to 25, which were already available with 4\,yr Pass 7 data. The TS map (Fig.~\ref{bzu0909}) confirms that the observed \gr\ signature emerges as a point-like source from a low-TS background.

\begin{figure}[]
   \centering
    \includegraphics[width=1.0\linewidth]{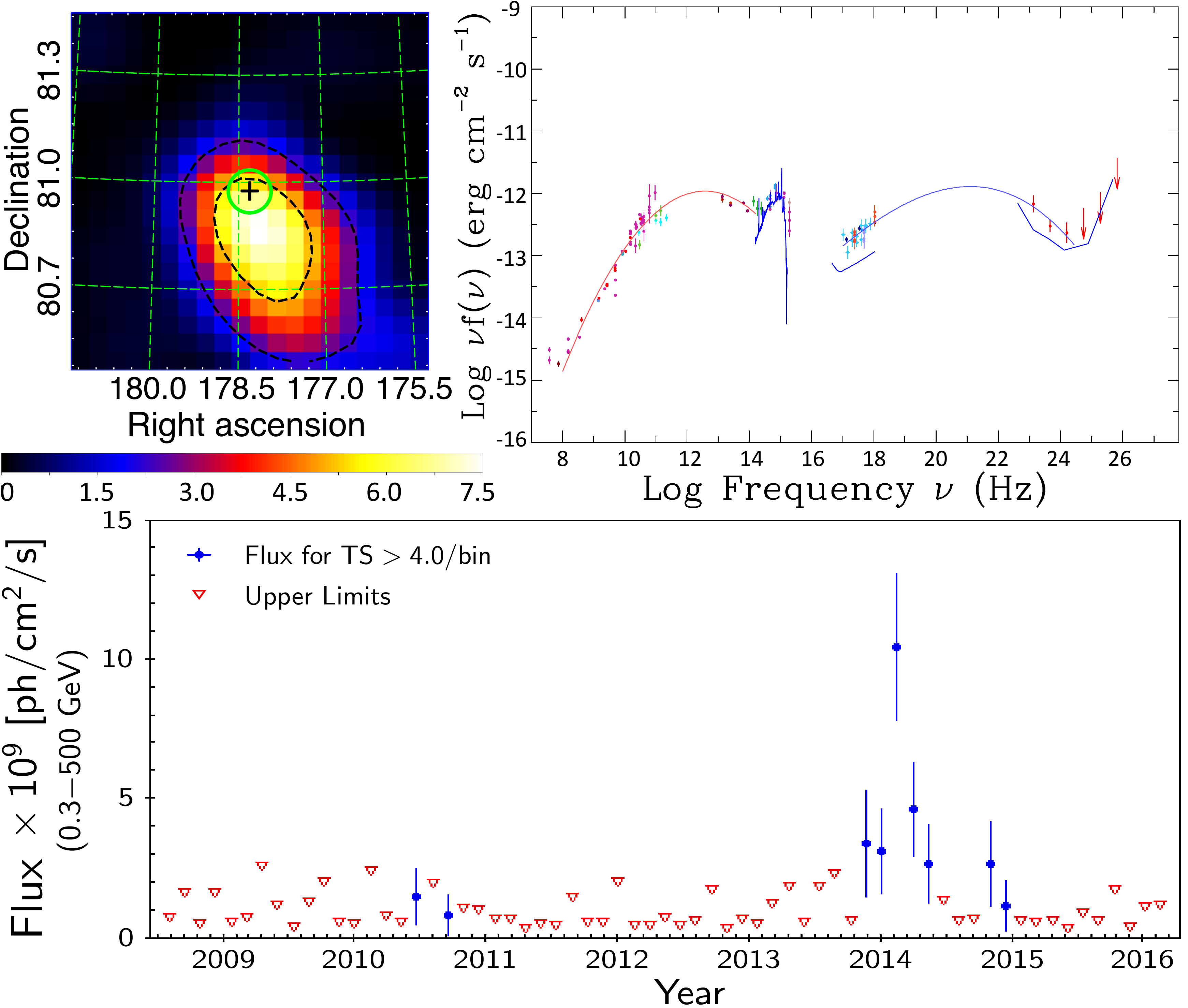}
     \caption{{\bf 5BZQJ\,1153+8058}. Top left: TS map using E\,$>$\,3.0\,GeV photons collected during the 2014 flaring period. Top right: SED for 5BZQJ\,1153+8058 also showing the template for thermal emission from accretion in the range 10$^{14}$ - 10$^{18}$\,Hz assuming z=1.250. Bottom: The \gr\ light curve for 5BZQJ\,1153+8058 along 7.5\,yrs.}
      \label{bzq1153}
\end{figure}

{\bf 5BZQ\,J1153+8058.} From the light curve (bottom panel of Fig.~\ref{bzq1153}) this source shows \gr\ activity only in 2014, which is consistent with its noninclusion in the 3FGL catalog (3FGL covers the period of 08/2008-08/2012). A likelihood analysis with 3FGL setup results in null detection. For this case, we are likely probing the peak of a transient \gr\ activity during the year 2014, and smoothing out the signal along 7.5 years of binned analysis.  The detection of a steady high-energy component (non-flaring state) is currently limited by the Fermi-LAT sensitivity. Therefore, the parameters reported on Table~\ref{detectiontab} embody the mean \gr\ spectrum behavior and represent a good example of how \gr\ activity is washed out for the building of current high-energy catalogs that integrate Fermi-LAT observations over 4.0 year (in case of the 3FGL) to 7.5 years. We built a TS map (top left in Fig.~\ref{bzq1153}) integrating along the whole year of 2014, showing that the blazar position is compatible with the \gr\ signature within the 68\% containment region. This case in particular allowed us to use E\,$>$\,3.0\,GeV photons instead of E\,$>$\,500\,MeV, providing improved localization. 

\begin{figure}[]
   \centering
    \includegraphics[width=1.0\linewidth]{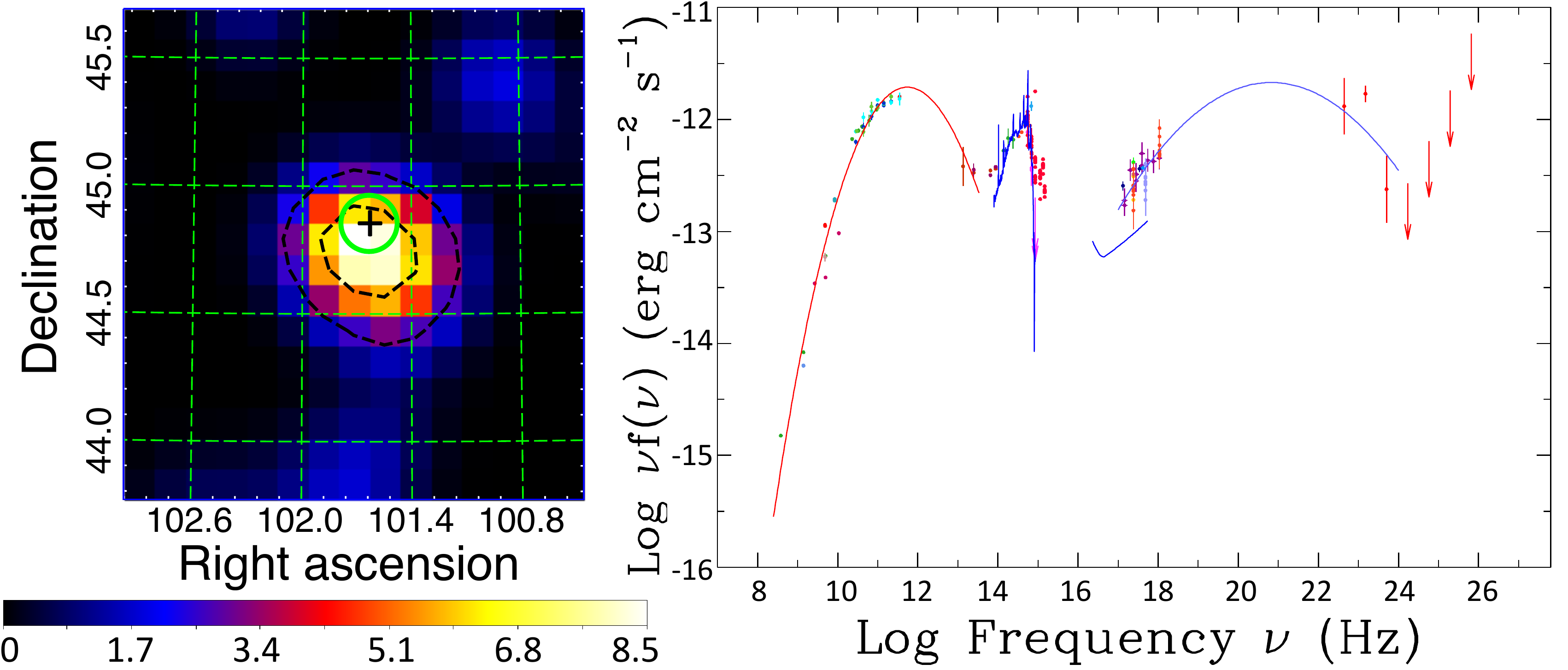}
     \caption{{\bf 5BZQJ\,0646+4451}. Left: TS map considering 500\,MeV to 12\,GeV photons and integrating over the flaring period from October 25 to November 30, 2010. Black dashed lines represent 68\% and 95\% confinement regions for the \gr\ signature. Right: SED for 5BZQJ\,0646+4451, also showing the blue bump template assuming z=3.396.}
      \label{bzq0646}
\end{figure}

{\bf 5BZQJ\,0646+4451.} Although this source is located in a relatively crowded region close to the Galactic plane ($|b|$=17.9$^{\circ}$), the low-energy detection is very significant with TS$^{Pass\,8}_{7.5\,yrs} \approx$\,80 in the 300-550\,MeV energy bin alone. We found no significant flaring activity during the 7.5\,yrs of observations with Fermi-LAT, and the period with the most significant \gr\ signature extends from October 25 to November 30, 2010. The highest energy photons detected from this region are $\sim$10\,GeV, and therefore we built a TS map in the 500\,MeV - 12\,GeV energy range (Fig.~\ref{bzq0646}, left panel). The \gr\ signature emerges as a point source fully compatible with the blazar position. In particular, the \gr\ counterpart has high redshift of z\,=\,3.396, and absorption due to EBL might hinder the detection of VHE photons. 

A likelihood analysis using the 3FGL setup has shown TS$^{Pass\,7}_{4\,yrs}$\,=\,20.3, with pivot energy E$_0$\,=\,1000\,MeV, pre-factor N$_0$\,=\,2.6\,$\times$\,10$^{-13}$\,ph/cm$^2$/s/MeV, and $\Gamma$\,=\,3.61, which is in good agreement with the parameters presented in Table~\ref{detectiontab}. This is another example of low-significance \gr\ signature with TS between 10 and 25, which would have been beneficial to report on without compromising the spectral description (that can be refined with longer integration time as shown).

\begin{figure}[]
   \centering
    \includegraphics[width=1.0\linewidth]{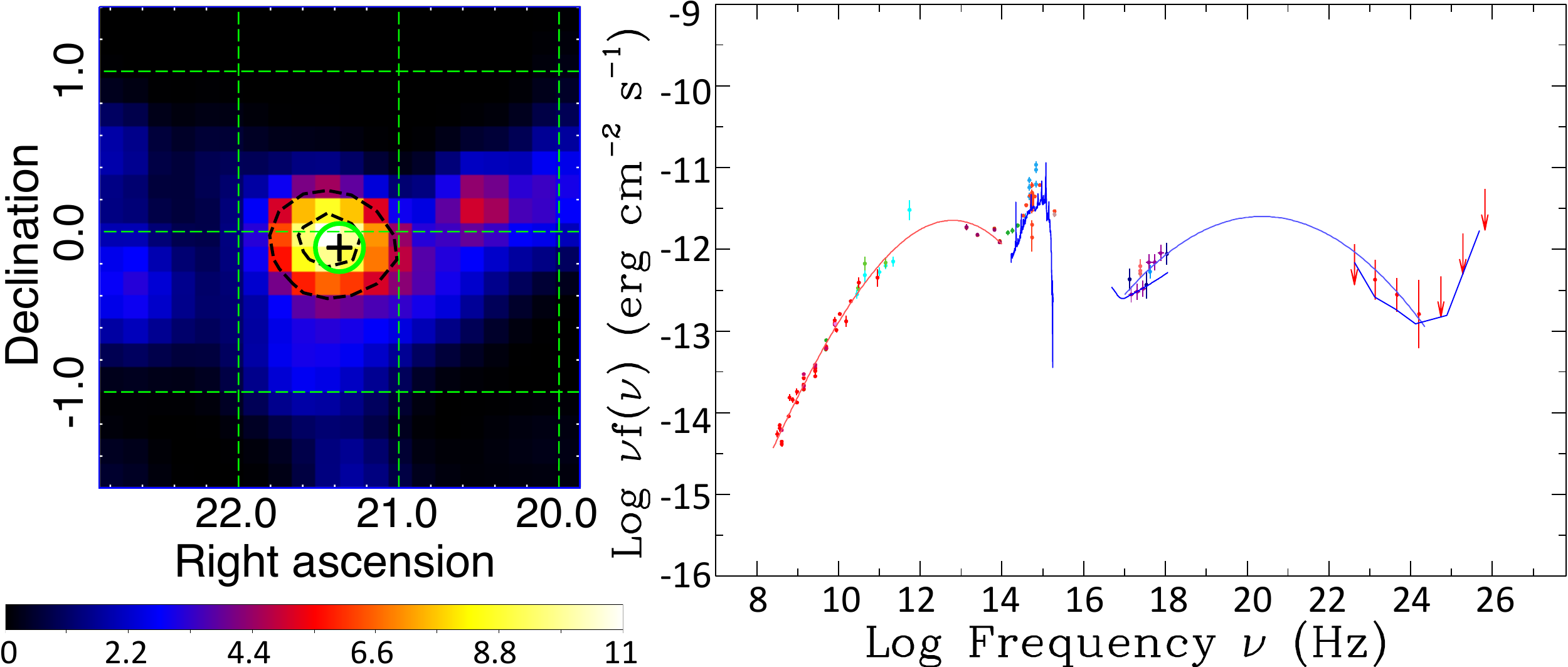}
     \caption{{\bf 5BZQJ\,0125-0005}. Left: TS map considering only E $>$ 500\,MeV photons and integrating data along the brightest month from July 31 to August 31, 2013. Black dashed lines show the 68\% and 95\% confinement region for the \gr\ signatures. Right: Multifrequency SED for 5BZQJ\,1153+8058 with the blue bump template corresponding to z=1.077}
      \label{bzq0125}
\end{figure}

{\bf 5BZQJ\,0125-0005.} This source has been detected with TS\,$\approx$\,20 when integrating over 7.5\,yrs with Pass 8 data. Its light curve does not present strong flaring episodes, but we identify the most relevant bin as July 31 to August 31, 2013, as used to build the TS map. Although the TS value is below 25, we consider it a firm detection since the TS map (\ref{bzq0125}) clearly shows the \gr\ point-like signature emerge from a low-TS background, and the 68\% confinement region is compatible with the blazar position. A likelihood analysis with the 3FGL setup gives TS$^{Pass\,7}_{4\,yrs}$\,=\,12.9 with parameters E$_0$\,=\,1000\,MeV, N$_0$\,=\,2.8\,$\times$\,10$^{-13}$\,ph/cm$^2$/s/MeV, and $\Gamma$\,=\,2.68, again in agreement with  parameters from Table~\ref{detectiontab}.   

\subsection{Source confusion}
\label{confusion}

\begin{figure}[]
   \centering
    \includegraphics[width=1.0\linewidth]{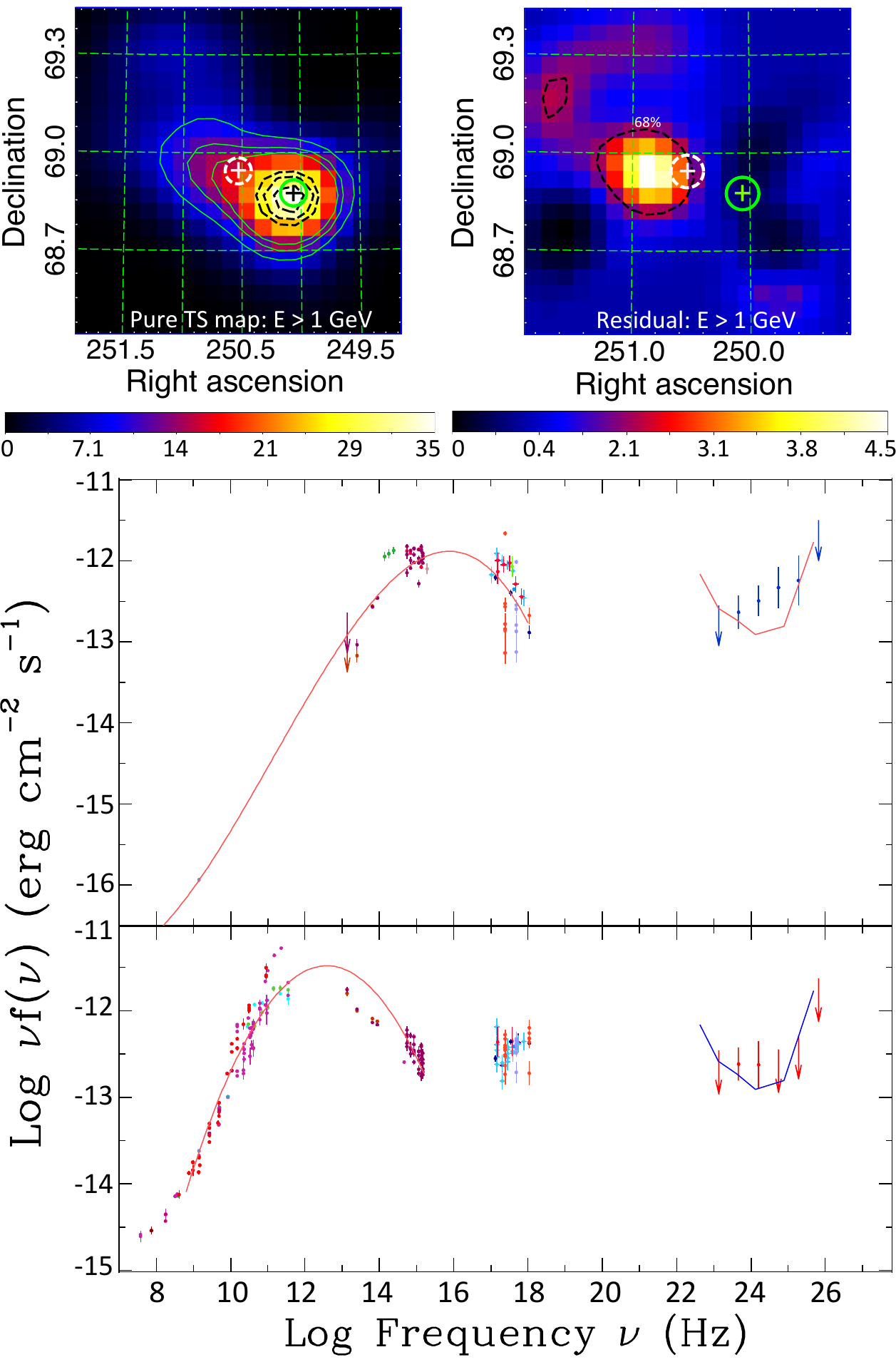}
     \caption{{\bf BZQJ\,1642+6856}. Top left: TS map for E\,$>$\,1.0\,GeV. The green thin lines show contours corresponding to TS values of 15, 12, 8, and black dashed lines represent 68\%, 95\%, and 99\% containment region. Top right: Residual TS map for E\,$>$\,1\,GeV considering a source placed at the 2WHSP position, indicated with green cross, as part of the \gr\ background; in this case the black dashed line corresponds to the 68\% confinement radius for the \gr\ source, while the white cross indicates the position of  BZQJ\,1642+6856. Middle and bottom: SEDs for 2WHSPJ\,1640+6852 for which no optical identification is available yet, and BZQJ\,1642+6856 with z\,=\,0.751. The upper limits on the \gr\ spectra have been calculated only for energy bins with low significance.}
      \label{2whsp1640}
\end{figure}

We discuss a case in which source confusion involving a steep and a hard spectrum \gr\, source is likely present. Even though no \gr\ source is reported within 1$^{\circ}$ of 5BZQJ\,1642+6856  in any of the FGL catalogs, our likelihood analysis finds a \gr\ signature matching the position of this source, which is located only $\sim$10\,arcmin from the HSP blazar 2WHSPJ\,164014.8+685233. Indeed, the high-energy E\,$>$\,1.0\,GeV TS map (Fig.~\ref{2whsp1640} in top left panel) reveals a dominant \gr\ signature coincident with the 2WHSP source, however extending toward the position of 5BZQJ\,1642+6856. We then consider photons in the 0.1-500\,GeV band and a likelihood function including two nearby point-like emitters, which gives us an estimate of the \gr\ spectral properties for each source as shown in Table~\ref{tableFermi2}. Despite the low statistical significance associated with BZQJ\,1642+6856 \gr\ signature (TS\,$<$\,10), we note that the photon spectral index $\Gamma$ estimated for the pair 5BZQ \& 2WHSP is consistent with the hypothesis of confusion between a steep and a hard component as expected for nearby LSP and HSP blazars.   

As a consistency test, we built an additional high-energy TS map (E\,$>$\,1\,GeV) by adding a point-like source at the 2WHSP position modeled as a power-law with the same parameters as from Table~\ref{solved1}, such that  the 2WHSP \gr\ signature is now part of the background model. As a result (Fig.~\ref{2whsp1640}, top right panel) we reveal a residual point-like signature consistent with 5BZQJ\,1642+6856 within the 68\% confinement radius. All together, it suggests this might be a case of source confusion, which is hard to resolve with currently available data. Since those \gr\ signatures have not been reported in previous high-energy catalogs, we present 2WHSPJ\,164014.8+685233 as a new detection and 5BZQJ\,1642+6856 as a relevant signature, whose resolved SEDs are shown in Fig.~\ref{2whsp1640} (middle and bottom panels).

\begin{table}[h]
\centering
\caption{Source model parameters derived from the Fermi Science Tools assuming a power-law to describe the \gr\ spectrum within the 0.1-500\,GeV energy band, with $N_0$ given in ph/cm$^{2}$/s/MeV, and assuming E$_0$ = 1000\,MeV.}  
\label{tableFermi2}
\begin{tabular}{c|ccc}
Source  & $N_0$ (10$^{-14}$) & $\Gamma$  & TS   
\\
\hline
2WHSPJ\,164014.8+685233   &  7.4$\pm$0.4  & 1.73$\pm$0.18 &  35.7  \\

5BZQJ\,1642+6856   &  9.1$\pm$4.0  & 2.31$\pm$0.28 &  8.3   \\
\end{tabular}
\label{solved1}
\end{table}

\subsection{New associations from improved positions determination}
\label{improveposition}

\cite{1BIGB} showed that high-energy TS maps can be used to improve the localization for many \gr\ signatures currently listed in FGL catalogs. It is well known that the Fermi-LAT detector \citep[][]{FermiLAT} is characterized by a highly energy-dependent point spread function (PSF), which contains 68\% of the 1\,GeV events within 0.8$^{\circ}$, decreasing afterward with a trend $\propto \, E^{-0.8}$ up 10's\,GeV, and finally roughly constant at 0.1$^{\circ}$ up to the highest energies considered in this paper. Therefore, working with E\,$>$\,1\,GeV allows us to better constrain the position associated with the \gr\ signature, which helps solve cases of source confusion. This is particularly important for unassociated 3FGL  \citep{FermiUnassociate1} that are actually counterparts of close by blazars. We present two such cases, noting that an approach based on the prior multifrequency identification of nearby blazars certainly improve the \gr\, associations with potential counterparts.

{\bf 3FGLJ\,0432.5+0539.} We present an improved position reconstruction for the source 3FGLJ\,0432.5+0539, for which no association is reported on the 3FGL catalog. We build a TS map using only photons with E\,$>$\,2\,GeV and removing the 3FGL source from the background model. 
\begin{figure}[h]
   \centering
    \includegraphics[width=1.0\linewidth]{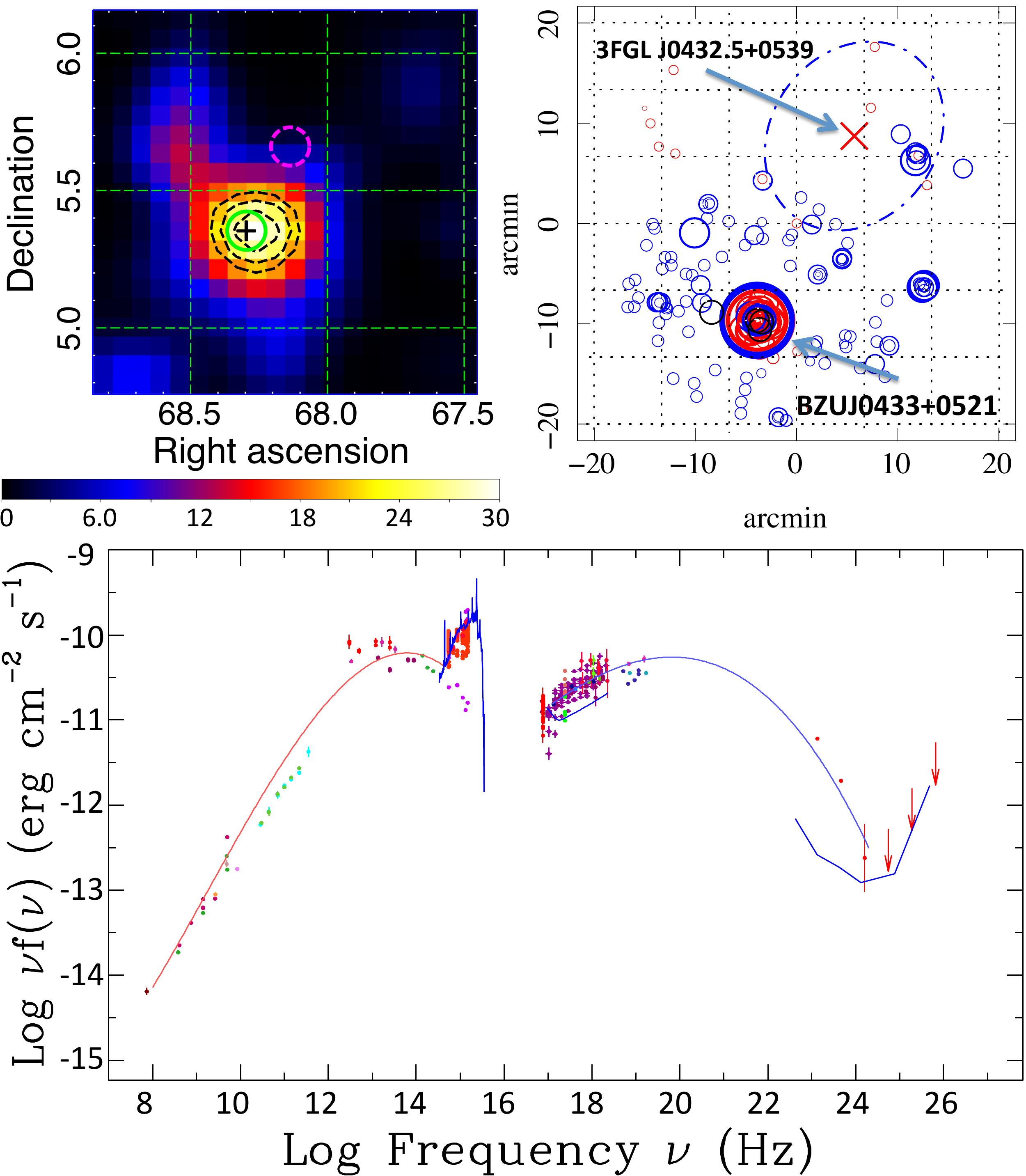}
     \caption{{\bf 5BZUJ\,0433+0521}. Top left: TS map built using only photons with E\,$>$\,2.0\,GeV; Black dashed lines show 68\%, 95\%, and 99\% confinement regions. The position of a nearby BZU source is highlighted by the green circle around the ``+" marker, while 3FGL detection is center on the magenta dashed circle. Top  right: an image obtained from the SSDC Sky-Explorer showing that 5BZUJ\,0433+0521 position is inconsistent with 3FGLJ\,0432.5+0539 detection also taking into account its 95\%  positional uncertainty (dot-dahsed ellipse). Blue and red circles represent respectively X-ray and radio frequency detections in the same region as taken from publicly available data. Bottom: SED for 5BZUJ\,0433+0521 with blue bump template  for z=0.033.}
      \label{bz0433}
\end{figure}

Fig.~\ref{bz0433} shows that the 68\% containment region for the \gr\ signature is fully consistent with the position of BZUJ\,0433+052, while the 3FGL position (magenta dashed circle on the top left panel) is well outside of the 99\% confinement region for the high-energy TS peak. Although part of the improvement could be ascribed to the better instrument response function (IRF) and event selection of the Pass 8 data release with respect to the Pass 7 as used for 3FGL production, it should be noted that high-energy TS maps has proven to provide a significant contribution in source positioning.

\begin{figure}[]
   \centering
 \includegraphics[width=1.0\linewidth]{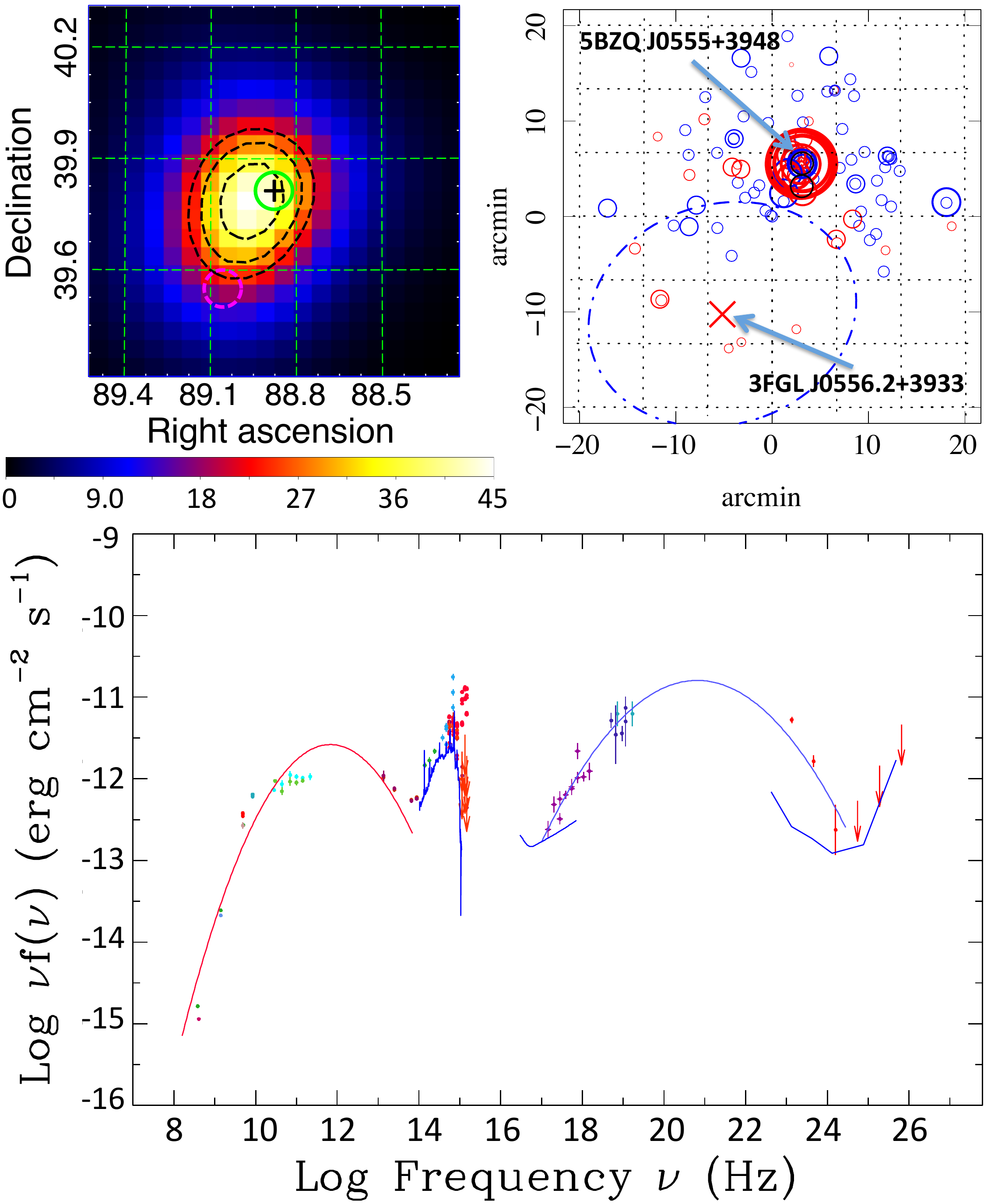}
     \caption{{\bf 5BZQJ\,0555+3948}. Top right: Sky-Explorer view, showing 3FGLJ\,0556.2+3933 indicated as a red cross, while 5BZQJ\,0555+3948 is outside the error circle associated with the \gr\ signature from 3FGL database. Blue and red circles represent X-ray and radio detections in this region. Top left: TS map, E\,$>$\,1.0\,GeV with improved position for the \gr\ signature; Black dashed lines show the 68\%, 95\%, and 99\% confinement region for the \gr\ signatures. The dashed magenta circle corresponds to the 3FGLJ\,0556.2+3933 position, while the green circle is centered on the position of 5BZQJ\,0555+3948 marked as black cross. Bottom: SED for 5BZQJ\,0555+3948, where the blue bump template assumes z=2.365.}
      \label{bz0555}
\end{figure}

{\bf 3FGLJ\,0556.2+3933.} This is another case of a 3FGL source with no association that benefits from an improved position reconstruction. As shown in the top right panel of Fig.~\ref{bz0555}, there is no relevant radio or X-ray counterpart that is compatible with the 95\% positional error ellipse for the 3FGL source (dot-dashed line). However, a powerful blazar is only $\sim$18 arcmin away. Indeed, building a TS map using high-energy photons (with E\,$>$\,1\,GeV) we are able to show that the 68\% confinement radius for the \gr\ signature is fully consistent with the position of 5BZQJ\,0555+3948. In addition, the \gr\ spectrum we obtained is compatible with expectations for the end tail of the IC bump.

\begin{figure}[h]
   \centering
    \includegraphics[width=1.0\linewidth]{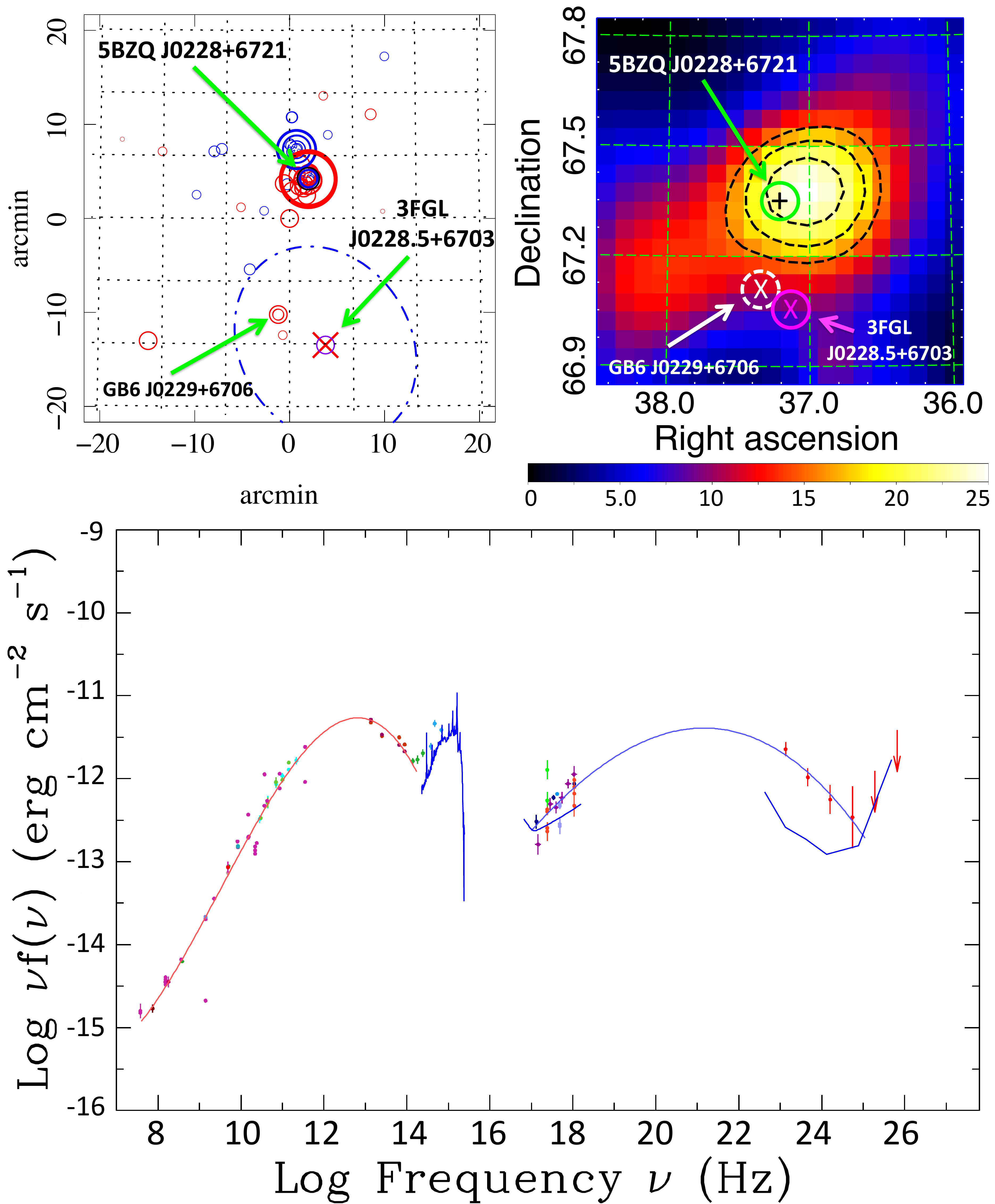}
     \caption{{\bf 5BZQJ\,0228+6721}. Top left: Sky Explorer view, showing 3FGLJ\,0228.5+6703 marked with ``x" symbol, and dashed line for the error circle associated with the 3FGL \gr\ signature. We also highlight the positions  of GB6 J0229+6706 and  5BZQJ\,0228+6721. The red and blue circles represent radio and X-ray detections in the field. Top right: TS map considering E\,$>$\,500 MeV  photons, with dashed lines representing 68\%, 95\%, and 99\% containment region for the \gr\ signature; 5BZQJ\,0228+6721 is center at ``+" matching the TS peak position. Bottom: SED for 5BZQJ\,0228+6721, with blue bump  assuming z=0.523.}
      \label{bzq0228}
\end{figure}

{\bf 3FGLJ\,0228.5+6703 and its correct counterpart.} Blazar 5BZQJ\,0228+6721 had no \gr\ counterpart in previous FGL catalogs, however a \gr\ source at its vicinity was detected (1FGLJ\,0233.4+6654) but with no consistent position. In this same region the source 3FGLJ\,0228.5+6703 has been associated with a radio source GB6J\,0229+6706 \citep[GB: from Green Bank 4.85 GHz northern sky survey,][]{GB6radio} which is within the 3FGL error circle (dot-dashed line, Fig. \ref{bzq0228} top left). This field is difficult to study because it is very close to the Galactic plane (at latitude b\,$\sim$\,6$^o$) and the intense low-energy diffuse \gr\ background can hinder both the source localization and detection when integrating over the full energy bandwidth.

Since the 3FGL position is only 5.94 arcmin from the blazar, we investigated whether we could improve the \gr\ localization of this source working with E\,$>$\,500 MeV photons, benefiting from lower background intensity, improved PSF, and longer integration time (from 4.0 to 7.5 years). We  calculated the E\,$>$\,500 MeV TS map (Fig. \ref{bzq0228}, top left) showing that the 68\% containment region for the \gr\ signature is  compatible with 5BZQJ\,0228+6721 (while GB6J\,0229+6706 is out of the 99\% containment region) and therefore the 3FGL association should be revised. We recalculated the \gr\ SED (Fig. \ref{bzq0228}, bottom) assuming a single source with position corresponding to 5BZQJ\,0228+6721, and found it is consistent with the end-tail IC bump.

\subsection{Low-significance \gr\ excesses}
\label{faintdetec}

We present sources showing faint \gr\ signature 10\,$<$\,TS\,$<$\,20 when integrating over 7.5 years of Fermi-LAT observations. It is important to report on faint detections, especially to clarify if these sources are actually \gr\ active, but under the TS limit currently used by the Fermi team, or if they are really quiet in the \gr\ band. Also, low-significance \gr\ detections help to complement the source number count in the faint end of the logN-logS$_{\gamma}$, and therefore can impact estimates of the contribution of blazars to the E\,$<$\,1 GeV extragalactic \gr\ background. We found the light curves for 5BZQJ\,2218$-$0335, 5BZBJ\,0006$-$0623, and 5BZQJ\,1038+0512  show short-lived activity at the timescale of a month, and we report on their TS maps and preliminary power-law modeling.

\begin{figure}[]
   \centering
       \includegraphics[width=1.0\linewidth]{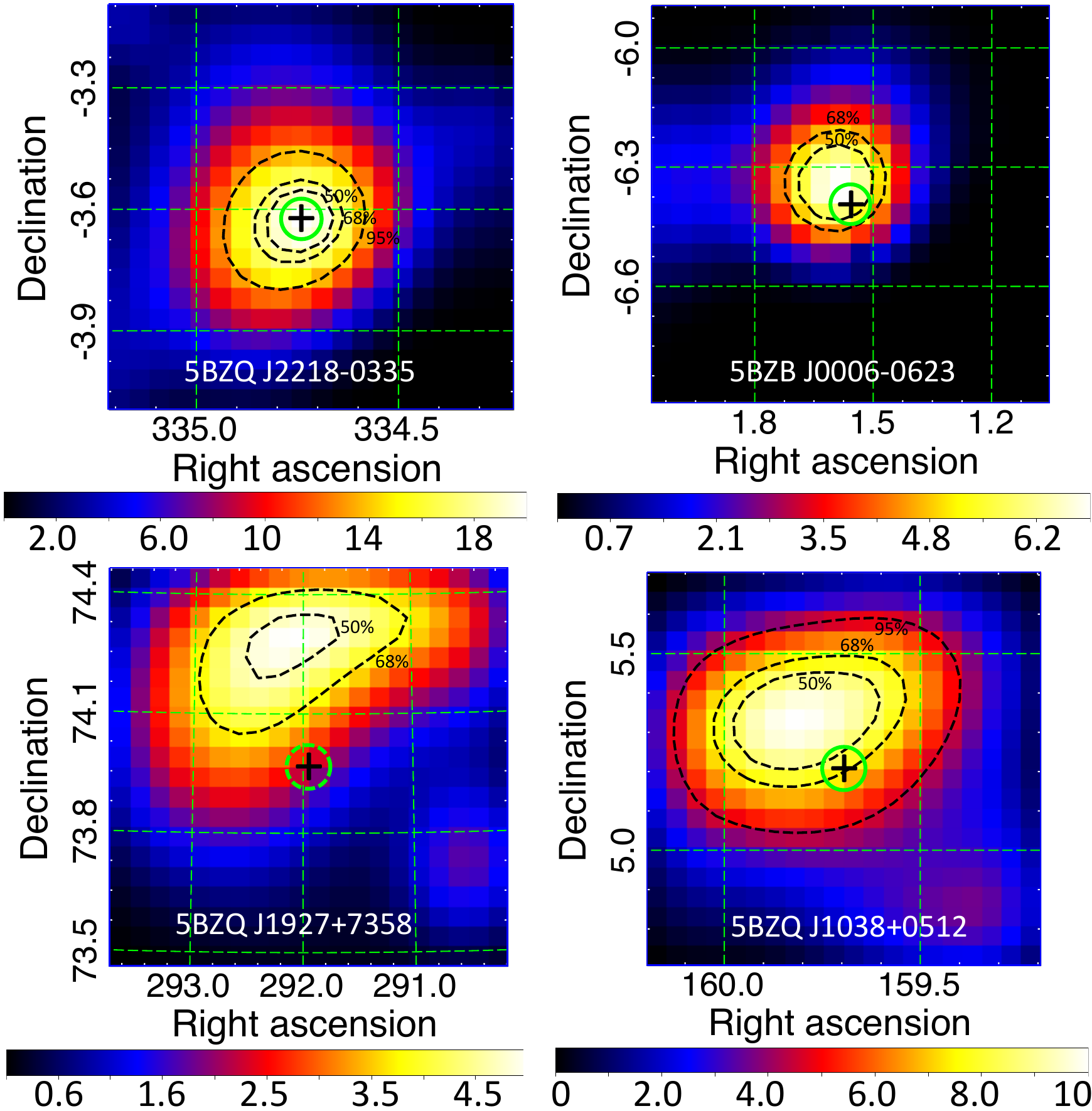}
    \caption{TS maps for the low-significance \gr\ detections using only E\,$>$\,500 MeV photons. For each case the blazar position is highlighted with thick green circle center at ``+". Black dashed lines show the 50\%, 68\%, and 95\% containment radius for the \gr\ signature.}
      \label{lowTS}
\end{figure}

\begin{table}[h]
\centering
\caption{Source model parameters from Fermi Science Tools, assuming a power-law to describe the \gr\ spectrum within 0.1-500\,GeV, with $N_0$ given in [ph/cm$^{2}$/s/MeV], and using E$_0$\,=\,1000 MeV as pivot energy. Case marked with * is probably an spurious detection.}  
\begin{tabular}{cc|ccc}
Source & z & $N_0$ (10$^{-13}$) & $\Gamma$   & TS   
\\
\hline
5BZQJ\,2218$-$0335 & 0.901 & 2.7$\pm$0.6   & 2.80$\pm$0.27 & 19.9  \\

5BZBJ\,0006$-$0623 & 0.347 & 1.4$\pm$0.4   & 2.13$\pm$0.20 & 17.5  \\

5BZQJ\,1927+7358* & 0.302 & 1.8$\pm$0.5   & 2.61$\pm$0.26 & 16.4  \\
 
5BZQJ\,1038+0512 & 0.45  & 1.6$\pm$0.6   & 2.79$\pm$0.38 & 10.1  \\

\end{tabular}
\label{fainttable}
\end{table}

As seen from the TS maps, only 5BZQJ\,1927+7358 is out of the 68\% containment for the \gr\ signature. In this case, the observed TS value (table \ref{fainttable}) when integrating over 7.5 yrs of observations can be attributed to residual signal from an unidentified close-by \gr\ source. For now, we consider it as a likely spurious signal.

\subsection{Detections during flaring episodes}
\label{flaring}

We present the \gr\ analysis for three radio-loud blazars that have relatively bright $\nu$f$_{\nu}$ Syn peak, 5BZQJ\,0010+1058, 5BZUJ\,0241-0815, and  5BZQJ\,2136+0041, for which the detection with Fermi-LAT was expected but not yet reported. This brings us to an important consideration about the role of short-lived flares at monthly timescales since a considerable portion of \gr\ active sources could be still undetected simply because their short-lived signatures are diluted below the sensitivity threshold when integrating Fermi-LAT observations over long exposure time.

\begin{figure}[]
\centering
    \includegraphics[width=1.0\linewidth]{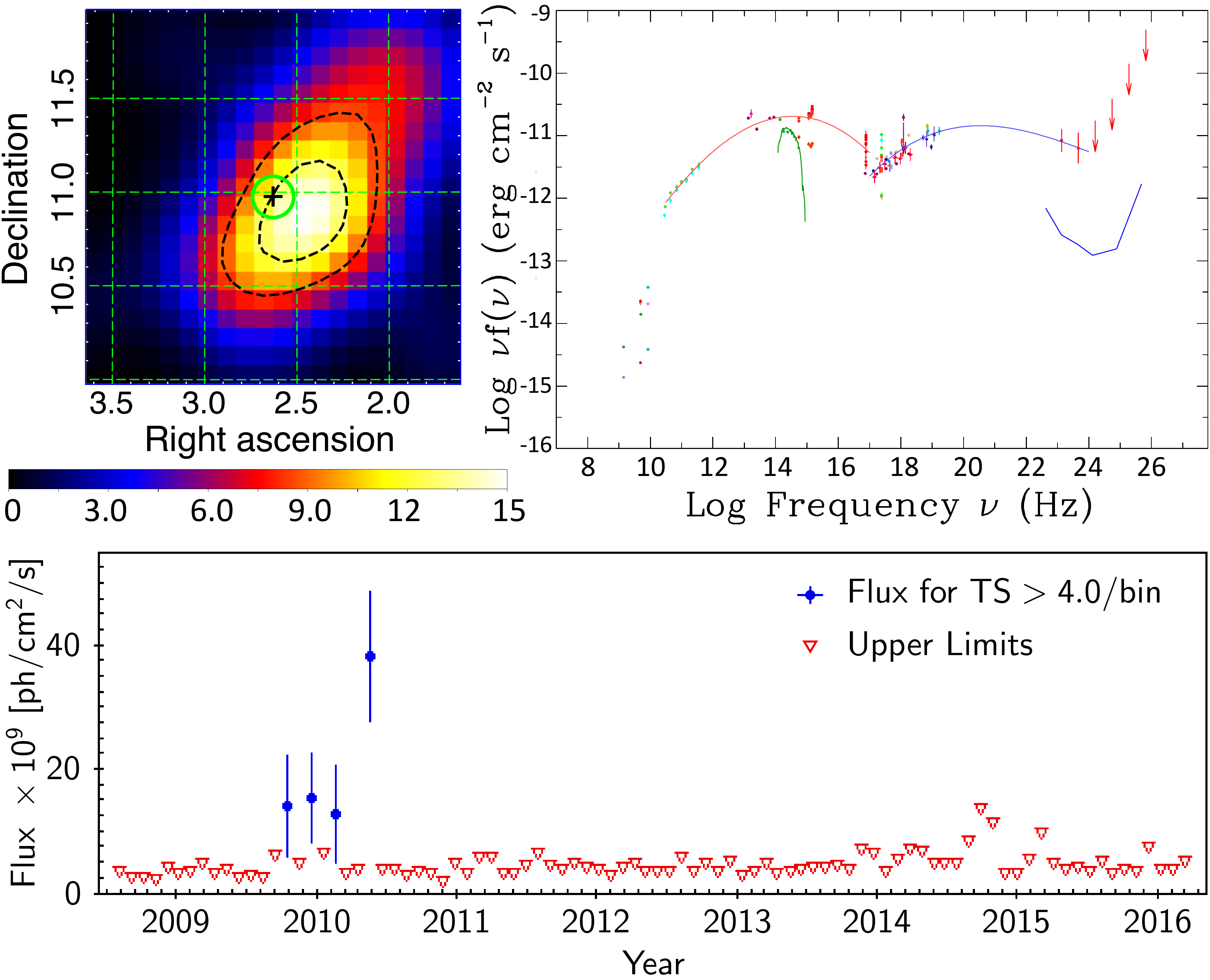}
     \caption{{\bf 5BZQJ0010+1058}. Top left: TS map considering E\,$>$\,500 MeV photons during a short-flare episode at MET: 296164808-298860398. Black dashed lines representing 68\% and 95\% containment region for the \gr\ signature. Top right: The corresponding SED of this object. Bottom: Light curve for 5BZQJ\,0010+1058 background extracted with likelihood analysis; flux points are calculated only for bins having TS\,$>$\,4 with a 30 day time bin along 7.5 yrs of observations, integrating 0.3-500\,GeV photons.}
      \label{bzq0010}
\end{figure}

{\bf 5BZQJ\,0010+1058 } This object is a radio-loud source (also known as MRK\,1501) for which \gr\, detection was expected since it is a relatively bright and close-by blazar with $\nu$f$_{\nu}$=10$^{-10.7}$ erg/cm$^2$/s and z=0.089. When integrating over 7.5 years, no \gr\, signature was evident (TS\,$\approx$\,0.0). Nevertheless its light curve (Fig. \ref{bzq0010}, bottom) shows fast (within the timescale of a month) and relatively bright flares, almost reaching two orders of magnitude variability with respect to the background. When running a likelihood analysis integrating only for the duration of the flare, June 18 2009 to October 31 2010, we could characterize the short-lived \gr\ spectrum, reaching TS\,$\approx$\,26 for the single month bin of June 2010. In this case (Fig. \ref{bzq0010}, top right), the upper limits calculated for the \gr\ SED are less restrictive, since we are integrating over a single month. We checked for coincident X-ray measurements from {\it Swift} (along the \gr\ flaring activity from June 2009 to May 2010), but for the observations made in February 2010 we could find no sign of strong X-ray flaring. This is a good example of transient \gr\ source, which we could only detect by means of a dedicated study focusing on blazars as \gr\ candidates. 

\begin{figure}[]
   \centering
    \includegraphics[width=1.0\linewidth]{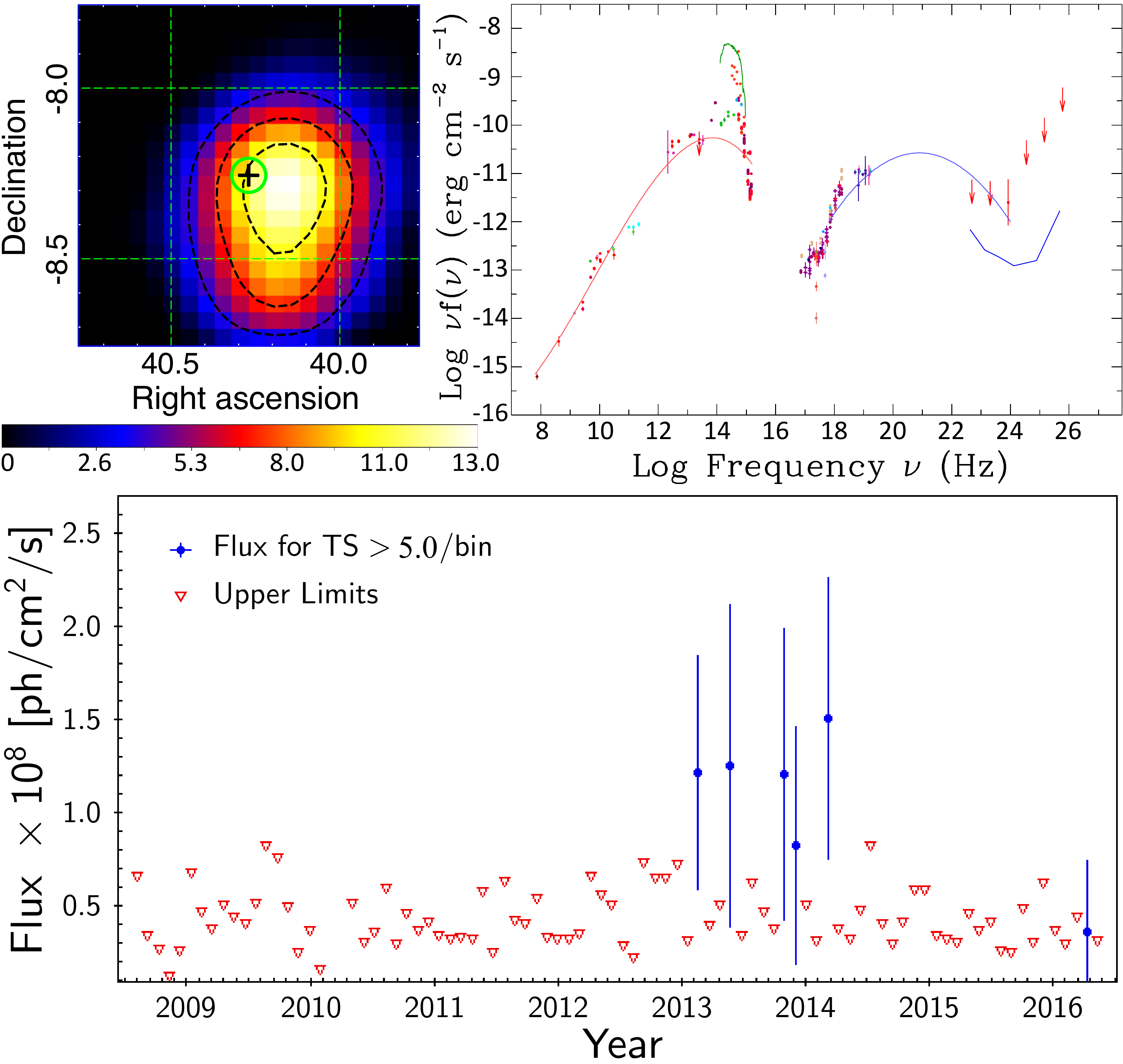}
     \caption{{\bf 5BZUJ\,0241-0815}. Top left: TS map for E\,$>$\,500\,MeV during a short-flare episode at MET: 415846402-418598890. The blazar position is highlighted by the green circle center at ``+"; the black dashed-lines are TS surfaces corresponding 68\%, 95\%, and 99\% confinement region for the \gr\ signature. Top right: The corresponding SED of this object; the green template corresponds to the elliptical galaxy emission at z=0.005. Bottom: Light curve for 5BZU\,J0241-0815. Flux points are only calculated for bins with TS\,$>$\,5 with a 30 day time bin along 7.5 yrs of observations, integrating 0.3-500\,GeV photons.}
      \label{bzu0241}
\end{figure}

{\bf NGC1052; 5BZUJ\,0241-0815.} In this case, the light curve shows a flaring state during 2013 to first quarter of 2014. From March 7 to April 7, 2014 the \gr\ signature reaches its highest state, at which TS\,$\approx$\,12. A TS map at E\,$>$\,500 MeV (fig. \ref{bzu0241}) shows that 5BZUJ\,0241-0815 is within the 68\% confinement region for the \gr\ signature. This blazar of unknown type is another example of a transient \gr\ emitter, which may contribute to building the currently unresolved \gr\ background. It is still not clear how to evaluate the impact of such sources for the observed extragalactic diffuse \gr\ component given that transient populations are not yet fully characterized.

\begin{figure}[]
   \centering
    \includegraphics[width=1.0\linewidth]{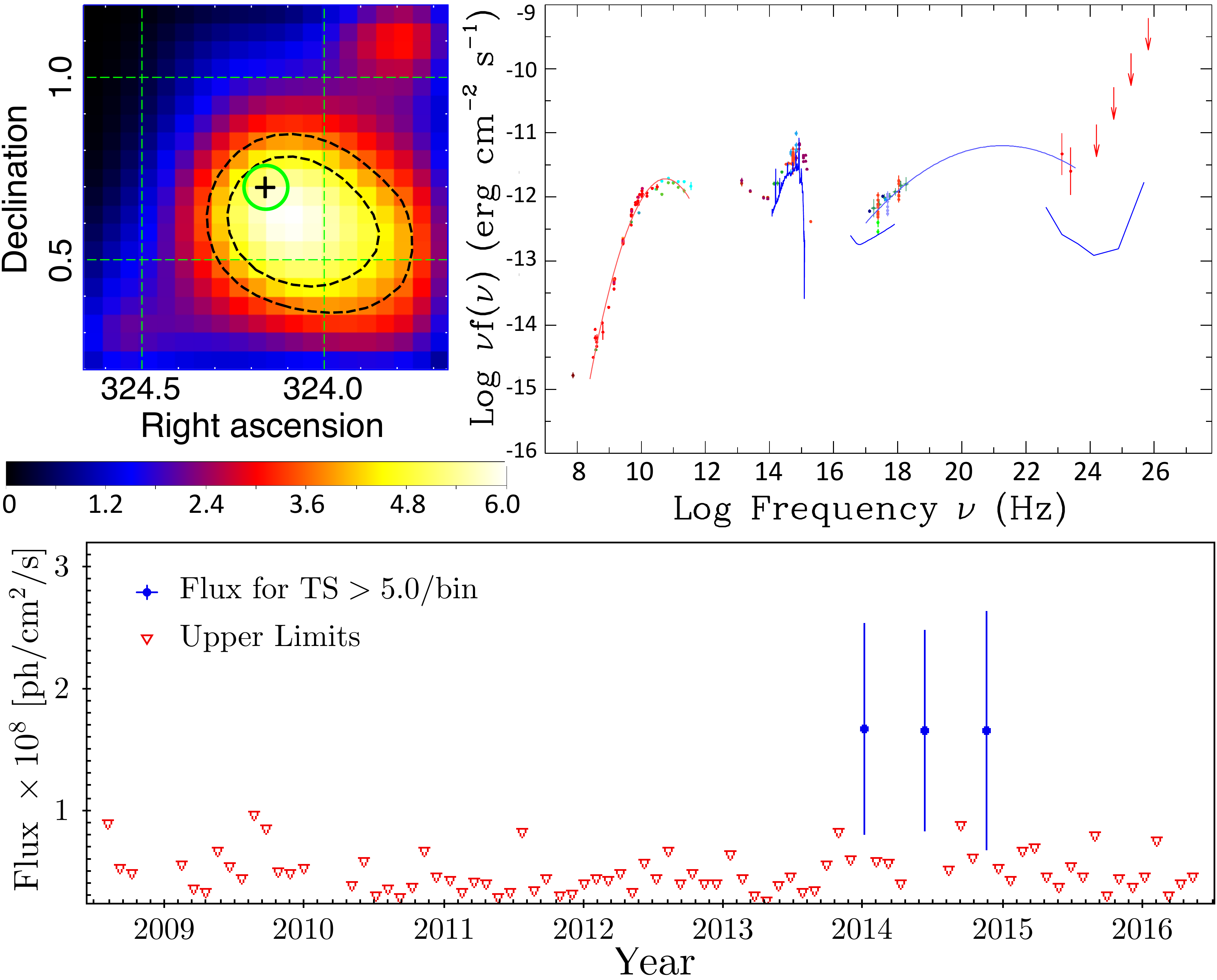}
     \caption{{\bf 5BZQJ\,2136+0041}. Top left: TS map for E\,$>$\,300MeV during a short-flare episode at MET: 410341424 413093913. The blazar position is highlighted by a green circle with center at ``+". The black dashed lines correspond to 50\% and 68\% confinement radius for the \gr\ signature. Top right: The corresponding SED of this object; the blue bump template assumes z=1.941. Bottom: Light curve for 5BZQJ2136+0041, flux is calculated only for TS\,$>$\,5 bins, with a 30 day time bin along 7.5yrs of observations, integrating 0.3-500\,GeV photons.}
      \label{bzq2136}
\end{figure}

{ \bf 5BZQJ\,2136+0041.} This source is classified as a FSRQ with z=1.941, and for this case we detected a short-lived \gr\ flare from January 2 to February 2, 2014, where TS$\approx$7.5. A TS map integrating over E$>$300 MeV photons (Fig. \ref{bzq2136}, top left) shows that the signature emerges as a point-like source and the 68\% containment region is compatible with 5BZQ J2136+0041 position. Although the \gr\ signature has low significance, it is important to report it as a potential faint \gr\ transient. Since we integrated along short time period, the lower limits computed in the \gr\ SED (Fig. \ref{bzq2136}, top right) are not so restrictive.

From the three cases presented in this section, two questions clearly arise:  How many AGNs could have short-lived flares, and how important is their integrated contribution to the \gr\ background probed by Fermi-LAT \citep{dgrbFermi} in the 100\,MeV to 800\,GeV energy range?

\subsection{Fermi-LAT nondetections}

We report on four Radio-Planck sources (5BZQJ 0927+3902, 5BZQJ\,2139+1423, 5BZQJ\,2022+6136 and 5BZQJ\,2007+4029) for which we could not find evidence of \gr\ signature during the 7.5 years of observations.

\begin{figure}[]
   \centering
    \includegraphics[width=1.0\linewidth]{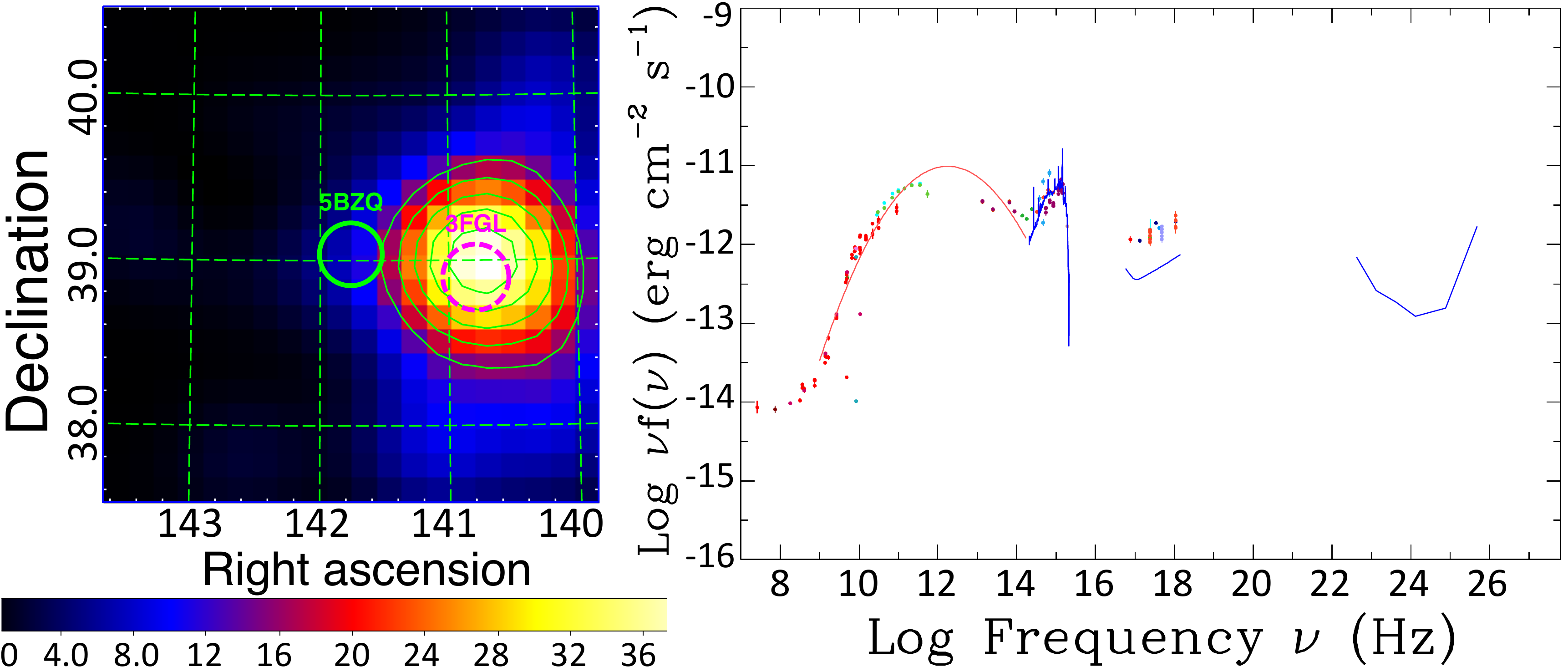}
     \caption{{\bf 5BZQJ\,0927+3902}. Left: TS map for 750-950 MeV photons. The blazar 5BZQJ\,0927+3902 is highlighted as a green circle and 3FGL is shown in magenta. Right: The SED with a polynomial fit to the mean Syn and the blue-bump template are represented for z=0.695. In the \gr\ band, the blue curve in the  10$^{22}$-10$^{26}$Hz band (0.1-500\,GeV) represents the Fermi-LAT four year sensitivity threshold, therefore an upper limit for the \gr\ emission.}
      \label{bzq0927ts}
\end{figure}

{\bf 5BZQJ\,0927+3902.} This is a bright radio blazar (z\,=\,0.695) about 46.1 arcminutes away from 3FGL J0923.1+3853 (which is associated to B2 0920+39). We investigate this region with a TS map, looking for signs of source confusion, but we could find none. The 3FGL \gr\ signature dominates the emission in this region as seen by the low-energy TS map (Fig. \ref{bzq0927ts}, left); the SED for 5BZQJ\,0927+3902 (Fig. \ref{bzq0927ts}, right) has no of \gr\ information. This is likely a good proxy for blazars with an IC component that is MeV peaked just as the following cases: 5BZQJ\,2139+1423, 5BZQJ\,2022+6136, and 5BZQJ\,2007+4029.

{\bf 5BZQJ\,2139+1423.} This blazar is classified as FSRQ, with relatively high redshift, z\,=\,2.427. A light curve with one month bins along 7.5 year showed no bins with TS\,$>$\,3.0. However, we do not discard \gr\ activity for this source. The IC component probably peaks at a frequency that is much lower than the bandwidth probed by Fermi-LAT, owing to its LSP frequency log($\nu_{peak}$)\,$\approx$\,11.0. In addition, because of its high redshift, the observed IC end tail gets even harder to probe. 

{\bf 5BZQJ\,2022+6136.} This source is also classified as FSRQ, although there is no good fitting between the optical to X-ray data and a blue-bump template (with z\,=\,0.228). There is a single episode from 27 September to 28 October 2011 for which a low significance $\gamma$-ray signal (TS\,=\,4.1) is present. 

{\bf 5BZQJ\,2007+4029.} This object is close to the Galactic plane, at latitude b\,=\,4.30$^{\circ}$, with relatively high redshift, i.e., z\,=\,1.736. A likelihood analysis integrating over 7.5 years, considering the full energy band 0.1-500\,GeV shows a very low-significance \gr\ signature, where TS\,$\approx$\,5.2, $\Gamma$\,=\,2.61$\pm$0.26, and N$_{0}$\,=\,2.4$\pm$1.2 $\times$ 10$^{13}$ ph/cm$^2$/s/MeV. However, from the TS map (300\,MeV-10\,GeV) there is no clear evidence for a point source, therefore we do not consider this as detection.


\section{Radio-Planck sample properties}

In previous sections, we showed that 99 of the 104 objects in the Radio-Planck sample have evidence for \gr\, emission at relevant level when integrating over 7.5 years of observation or during flaring episodes. By fitting the Syn and IC components with a third order polynomial \citep{giommisimultaneous}, we estimate their SED peak parameters (Table \ref{tableRadioPlanck}) using all available nonsimultaneous data. In particular, the \gr\ points are from the 3FGL catalog in the case of previously detected sources, and from our own data reduction in the case of newly detected sources, considering 7.5 years of Fermi-LAT observations; we also study statistical properties related to the nonthermal peak parameters.

\subsection{Synchrotron and IC peak parameters}

\begin{figure}[]
   \centering
    \includegraphics[width=1.0\linewidth]{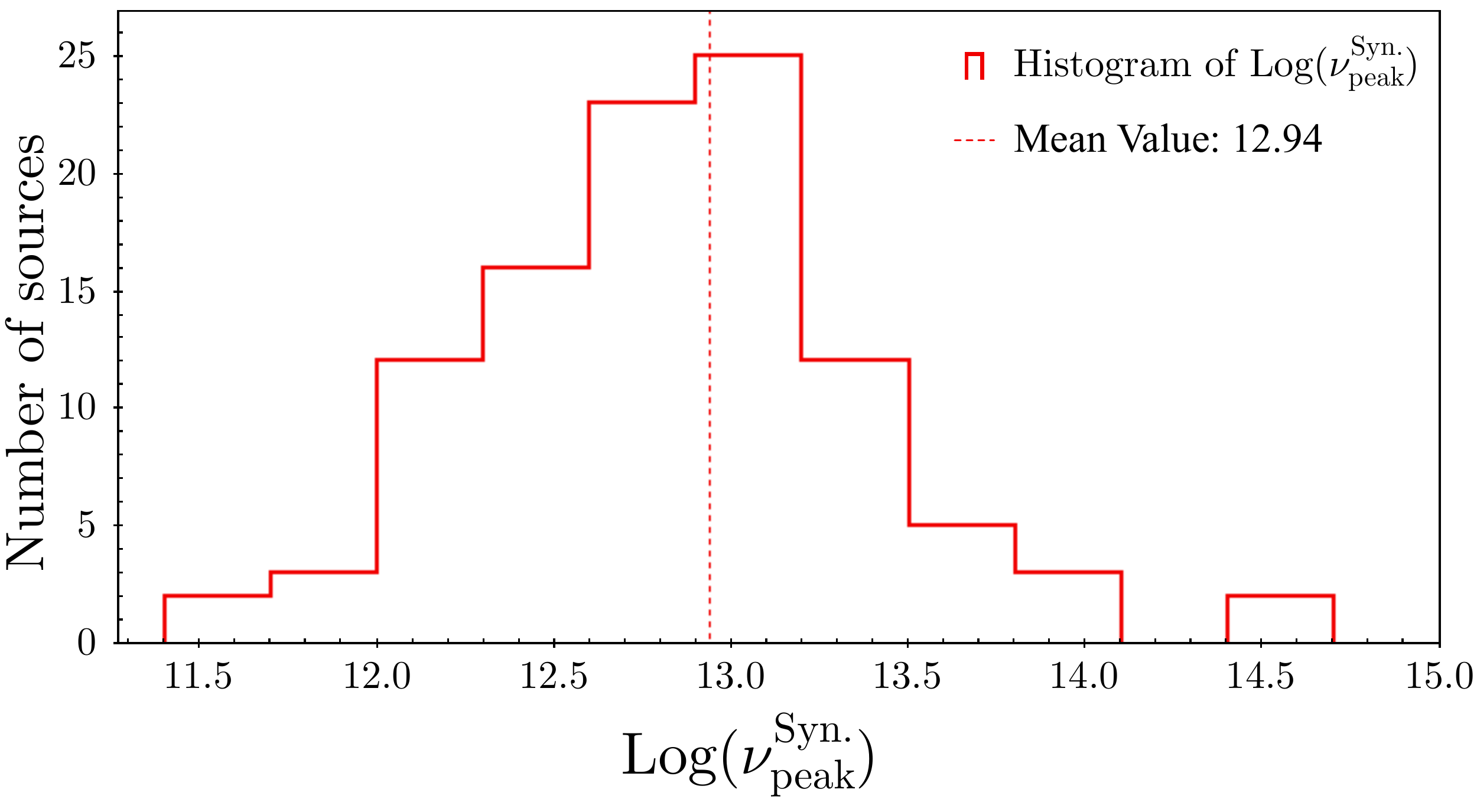}
     \caption{Distribution of log($\nu^{Syn}_{peak}$) [Hz] for the Radio-Planck sample considering all 104 objects, bin size of 0.3; the mean value of 10$^{12.94}$ Hz is shown as a dashed line.}
      \label{nudist}
\end{figure}

In Fig. \ref{nudist} we plot the distribution of log($\nu^{Syn}_{peak}$), showing that the Radio-Planck sample is dominated by LSP blazars, with mean value $\langle$log($\nu^{Syn}_{peak}$)$\rangle$\,=\,12.94$\pm$0.07\footnote{From here on, the errors to the mean values are calculated as $ \rm  \sigma/\sqrt{n}$ , where $\sigma$ is the standard deviation and n is the total number of objects in each sample (or subsample).} [Hz]. This is because our sample is flux limited in the microwave band where blazars of the LSP type are by far the most abundant objects.

\begin{figure}[]
   \centering
    \includegraphics[width=1.0\linewidth]{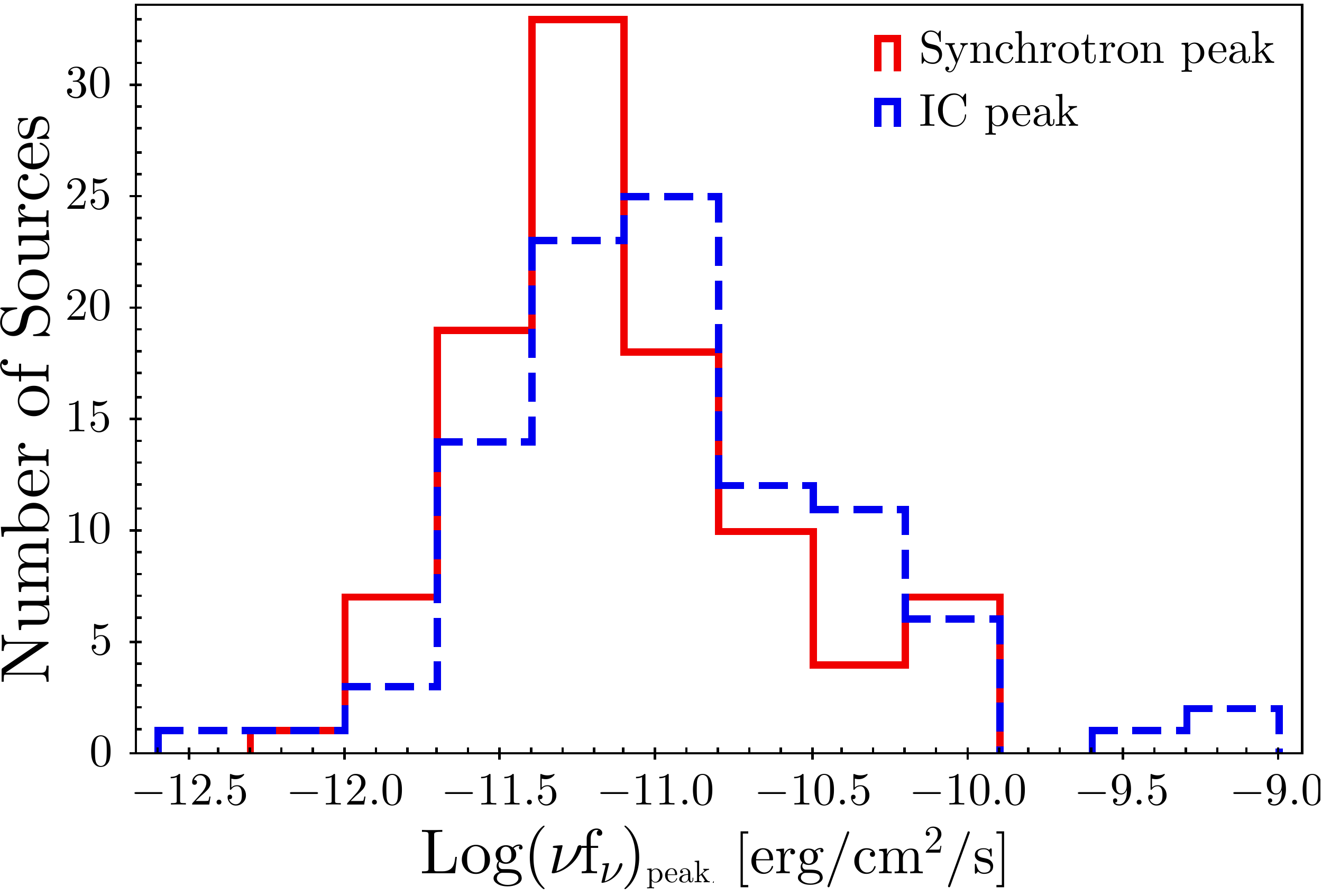}
     \caption{Histogram of log($\nu$f$_{\nu}$) at the Syn peak (in red) and the IC peak (in blue), considering all 99 sources in the Radio-Planck sample for which we could estimate the Syn and IC peak parameters.}
      \label{nfndist}
\end{figure}

In Fig. \ref{nfndist} we plot a histogram showing the distribution of log($\nu$f$_{\nu}$) peak values for both Syn and IC bumps. A parametric Kolmogorov Smirnov (KS) test comparing both histograms gives a p$_{value}$\,=\,0.86, implying that the luminosity distributions of both components are similar. In fact, the mean values of the peak fluxes are relatively close, i.e., $\langle$log($\nu$f$^{Syn}_{\nu}$)$\rangle$\,=\,-11.11$\pm$0.05 [erg/cm$^2$/s]; $\langle$log($\nu$f$^{IC}_{\nu}$)$\rangle$\,=\,-10.94$\pm$0.05 [erg/cm$^2$/s]. This similarity in the peak power distribution of the two components suggests that on average the ratio of $\nu$f$^{Syn}_{\nu}$\, to \,$\nu$f$^{IC}_{\nu}$ values might be close to one for the population of LSP blazars.

\subsection{Peak ratio parameter}

\begin{figure}[]
   \centering
    \includegraphics[width=1.0\linewidth]{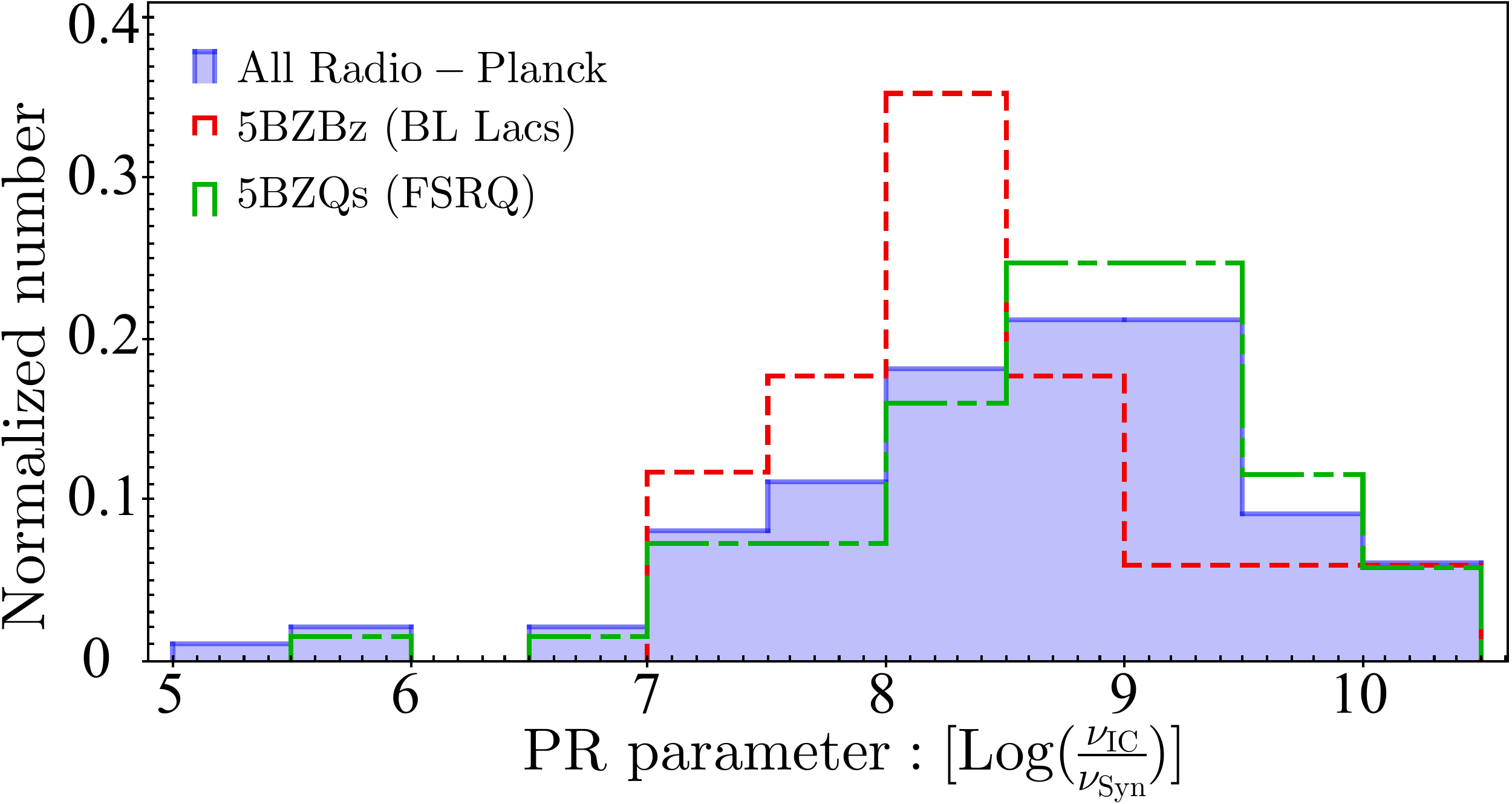}
     \caption{Peak ratio log($\rm \nu^{IC}_{peak} / \nu^{Syn}_{peak}$) distribution. Full indigo bars represent the whole Radio-Planck sample, red dashed bars indicated the subsample of BL Lacs, and green dot-dashed bars show the subsample of FSRQ.}
      \label{nuRatio}
\end{figure}

The $\nu^{Syn}_{peak}$ and $\nu^{IC}_{peak}$ parameters, which are determined in the log($\nu$f$_{\nu}$) versus log($\nu$) plane, are very representative of blazar SED properties because they give the peak energy where most of the Syn and IC power are emitted. To study their statistical properties we define the peak ratio (PR) parameter as the logarithm of $\nu_{peak}$ ratios: $\rm PR = log ( \nu^{IC}_{peak}/\nu^{Syn}_{peak} )$
and plot its distribution (Fig. \ref{nuRatio}). We determine the characteristic mean value for PR when considering all LSP blazars: $\langle$PR$_{all} \rangle$=8.60$\pm$0.09. We also show the distribution of PR values for the subsamples classified as 5BZB (BL Lacs), where the mean value $\langle$PR$_{BL \ Lacs} \rangle$=8.42$\pm$0.20 and 5BZQs (FSRQ), where the mean value $\langle$PR$_{FSRQ} \rangle$=8.75$\pm$0.09. There are no significant differences for the average PR parameters, since variations are contained within errors. 

\begin{figure}[]
   \centering
    \includegraphics[width=1.0\linewidth]{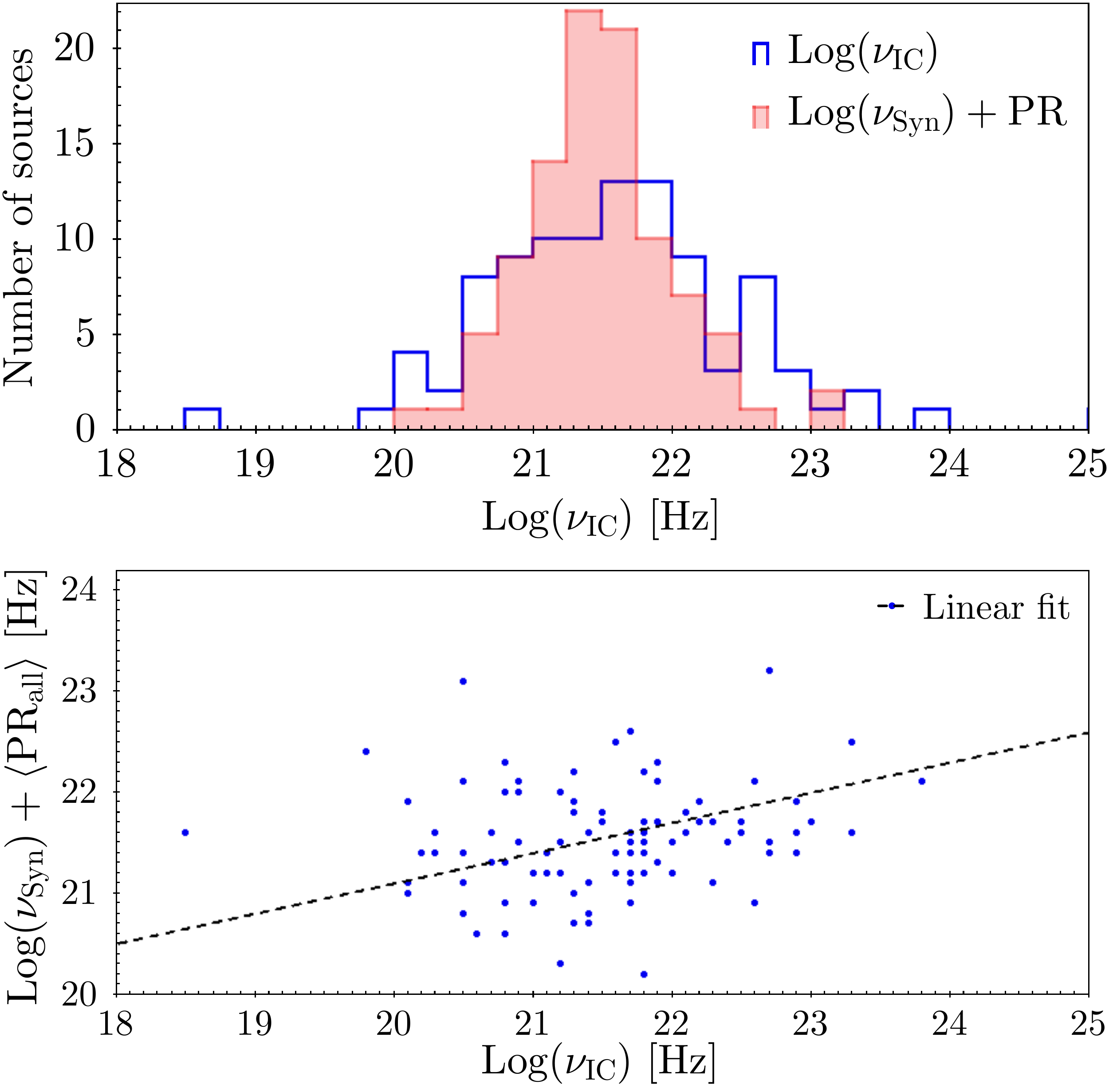}
     \caption{Top: Distribution of log($\nu^{IC}_{peak}$) in blue, using measured values from fitting the IC component, and in pink using log($\nu^{IC}_{peak}$) = log($\nu^{Syn}_{peak}$)+$\langle$PR$_{all} \rangle $ to estimate $\nu^{IC}$ based on $\nu^{Syn}$ measurements. Bottom: A scatter plot with log($\nu_{IC}$) values vs. log($\nu_{Syn}$)+$\langle$PR$_{all}\rangle$ together with a linear fit showing week correlation.}
      \label{PeakPrediction}
\end{figure}

In Fig. \ref{PeakPrediction} we show the log($\nu^{IC}_{peak}$) distribution for all 98 LSPs with available IC data (in blue), which has mean value of $\langle$log($\nu_{IC}$)$\rangle$=21.53$\pm$0.90 [Hz]. We then use the $\langle$PR$_{all} \rangle$ value to estimate $\nu^{IC}_{peak}$ based on the $\nu^{Syn}_{peak}$ according to log($\nu^{IC}_{peak}$) = log($\nu^{Syn}_{peak}$)+$\langle$PR$_{all} \rangle $. The distribution of $\nu^{IC}_{peak}$ calculated via $\langle$PR$_{all} \rangle$ parameter is shown in pink, which is indeed well described by this simple relation. Most likely, there is a dominant process connecting Syn and IC bumps, otherwise such correlations would not show up. If multiple emission scenarios were at work, we would expect large spreading in the parameter space, not tight Gaussian distributions. As known, the most established picture to describe the SED shape assumes dominant SSC leptonic scenario, but there is extensive discussion in the literature considering the role of the EC for different blazars, even reporting on observable evidence \citep{EC-evidence}.

A scatter plot with log($\nu_{IC}$) values versus log($\nu_{Syn}$)+$\langle$PR$_{all}\rangle$ shows that there is only marginal evidence for the correlation between those parameter (Fig. \ref{PeakPrediction}, bottom) given that the Pearson's correlation coefficient {\it r} for a linear fit is $\sim$0.17. The fact that the PR parameter helped to describe the distribution of log($\nu_{IC}$) as a population, tells us that the log($\nu_{IC}$) measured by our fitting might have large uncertainties; given the absence of a significant correlation, when comparing case by case with the scatter plot. Indeed there is a large uncertainty for that parameter mainly due to the huge data gap in the energy window between tens of KeV up to hundreds of MeV, which is smoothed out when considering the whole population of bright LSP sources.

\subsection{Compton dominance}

The ratio of IC to Syn peak power is known as the Compton dominance (CD). This is an important parameter for describing blazar SEDs \citep{CompDom,CDblazars,CDsequency2017A&A}, since it measures the dominant power output component for each source, i.e., 
 
\begin{equation}
\rm CD =  \frac{L_{IC}}{L_{Syn}}  \equiv \frac{\nu f^{IC}_{\nu}}{\nu f^{Syn}_{\nu} }  \ ; at \ the \ SED \ peaks
.\end{equation}

Luminosity is written as L\,=\,4$\pi$d$^2_L$ $\nu$f$_{\nu}$/(1+z)$^{1-\alpha}$, where d$_L$ represents the luminosity distance and the (1+z)$^{1-\alpha}$ factor is the k-correction assuming a power-law spectrum with energy index $\alpha$. Since we calculate the CD using the luminosity ratio at the peak power,  $\alpha$=1 for both IC and Syn peaks, and the luminosity ratio is simply the flux ratio.

\begin{figure}[]
   \centering
    \includegraphics[width=1.0\linewidth]{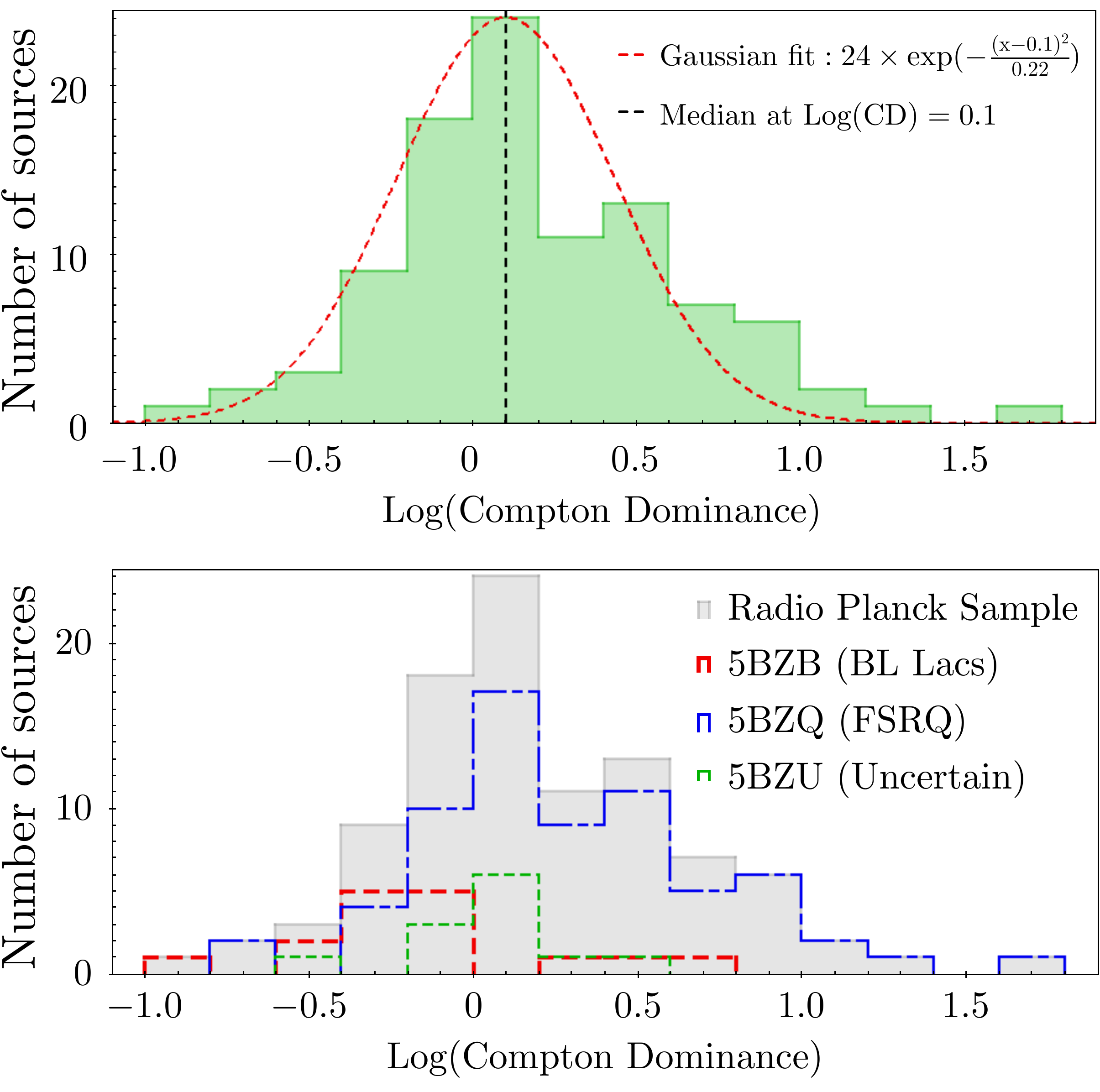}
     \caption{Log(CD) distribution for the Radio-Planck sample. Top: The green bars represent all 99 cases that have Syn + IC data available for calculating the CD parameter. The red dashed line represents a Gaussian function with $\sigma^2$=0.22 around the median value of log(CD)\,=\,0.1. Bottom: The log(CD) distribution for BL Lacs (red dashed lines), FSRQ (blue dot-dashed lines), and blazars classified as uncertain are shown in green dashed lines.}
      \label{dominance}
\end{figure}

In Fig. \ref{dominance} (top) we plot in green the distribution of  log(CD) parameter for the Radio-Planck sample, which has a median value of 0.1. The median is only slightly larger than 0, implying that on average, the peak-power output for the sync and IC components are similar. In Fig. \ref{dominance} (top), we add a tentative Gaussian fit to the log(CD) distributions, showing that a single Gaussian function (red dashed line) hardly describes the overall shape, particularly the highest CD values. The tail toward log(CD)\,$>$\,1.0 probably owing to strong variability in $\gamma$-rays compared to radio bands, pushing log(CD) to high values. The mean value $\rm \langle log(CD) \rangle$\,=\,0.17$\pm$0.05 is clearly affected by that, therefore a median might be more reliable as the representation of steady $\gamma$-ray activity in LSP blazars.

The histograms at bottom panel of Fig. \ref{dominance} represent the CD for different subsamples defined according to the 5BZcat classification, that is, BL Lacs, FSRQ, and unclassified sources. \cite{CDsequency2017A&A} have also estimated the CD for FSRQ and BL Lacs, but based on luminosities at fixed energies: $\rm L_{(1\,GeV)}$ as measured with Fermi-LAT, and $\rm L_{(3.4 \mu m)}$ as measured with the W1 channel from WISE satellite, for instance, $\rm CD = L_{(1\,GeV)}/L_{(3.4 \mu m)}$. Our measurements instead are taken at the peak of Syn and IC components. In both cases, there is a trend for BL Lacs to populate the log(CD) range with the lowest values (with mean $\rm \langle log(CD)_{(BL-Lac)} \rangle = -0.16 \pm 0.08$), while FSRQ populate a wider region (with mean $\rm \langle log(CD)_{(FSRQ)} \rangle = 0.28 \pm 0.06$ for the Radio-Planck sample). We also checked whether the log(CD) distribution depends on radio flux density splitting the sample in two subsets with $\rm f_{5GHz}$\,$>$\,1.5\,Jy and $\leq$\,1.5\,Jy, at 5\,GHz. As result, the two histograms turned out to be very similar (a KS test gives a p-value\,=\,0.693). Also a scatter plot of radio flux versus log(CD) shows no evidence of correlation between these quantities.

\begin{figure}[h]
   \centering
    \includegraphics[width=1.0\linewidth]{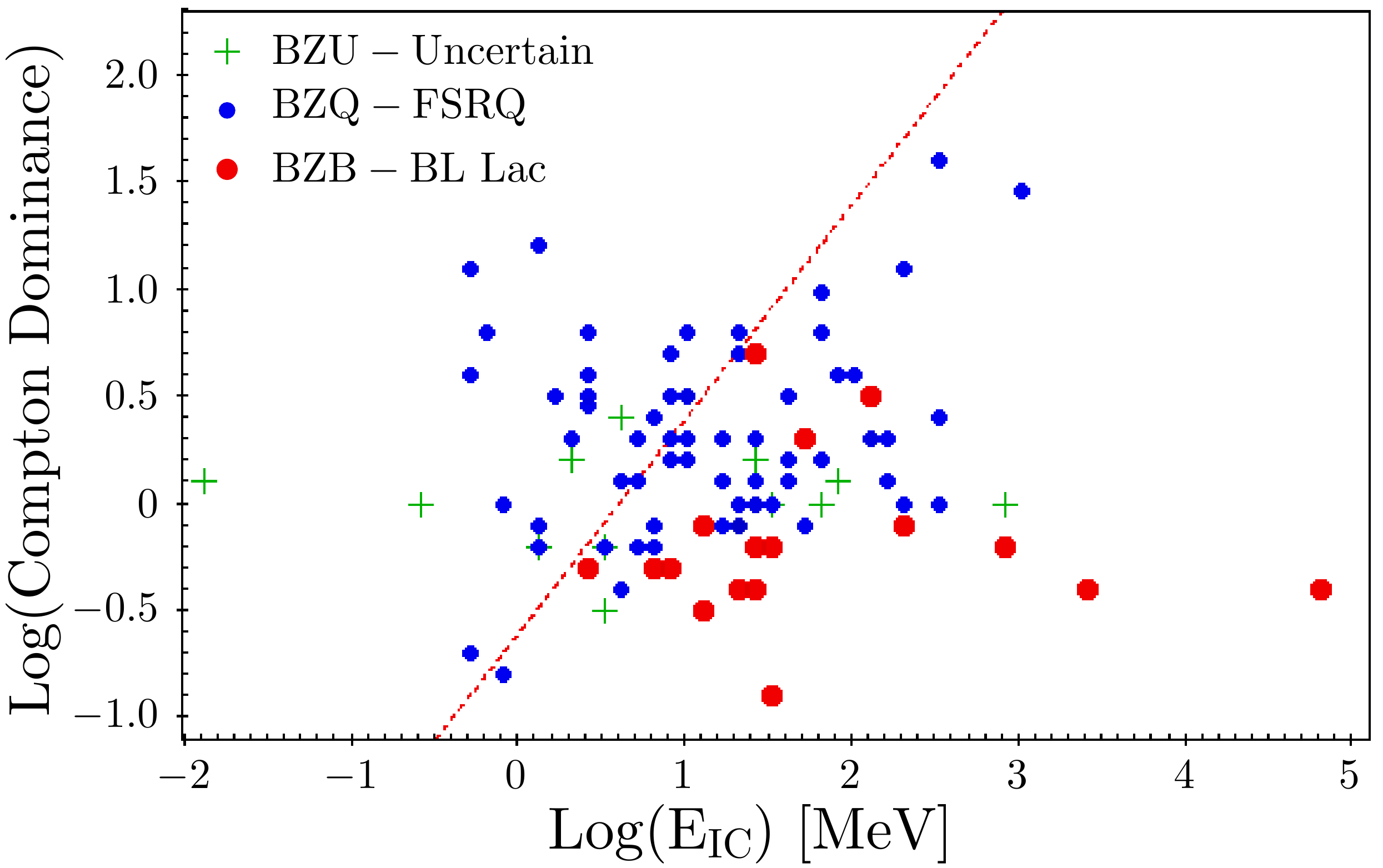}
     \caption{Compton dominance vs. energy associated with the IC peak ($\rm E_{IC}$) for the Radio-Planck sample. We plot FSRQs as blue dots, BL Lacs as red dots, and uncertain blazars as green crosses. The red dotted line sets a qualitative cut to highlight the region populated by BL Lacs.}
      \label{CDenergy}
\end{figure}

In Fig. \ref{CDenergy} we plot the CD versus IC peak energy ($\rm E_{IC}$, in MeV). When considering the whole sample, there is no clear correlation between those parameters (the Pearson's correlation coefficient $r \approx$\,0.13). However for individual classes (BL Lacs, FSRQ, and uncertain types) the BL Lacs tend to be associated with the highest IC peak energies, even though they are not dominating the end tail with high $\rm \log(CD)$ values (Fig. \ref{dominance}, bottom).

\subsection{Influence of variability on the Compton dominance}

\begin{figure}[h]
   \centering
    \includegraphics[width=1.0\linewidth]{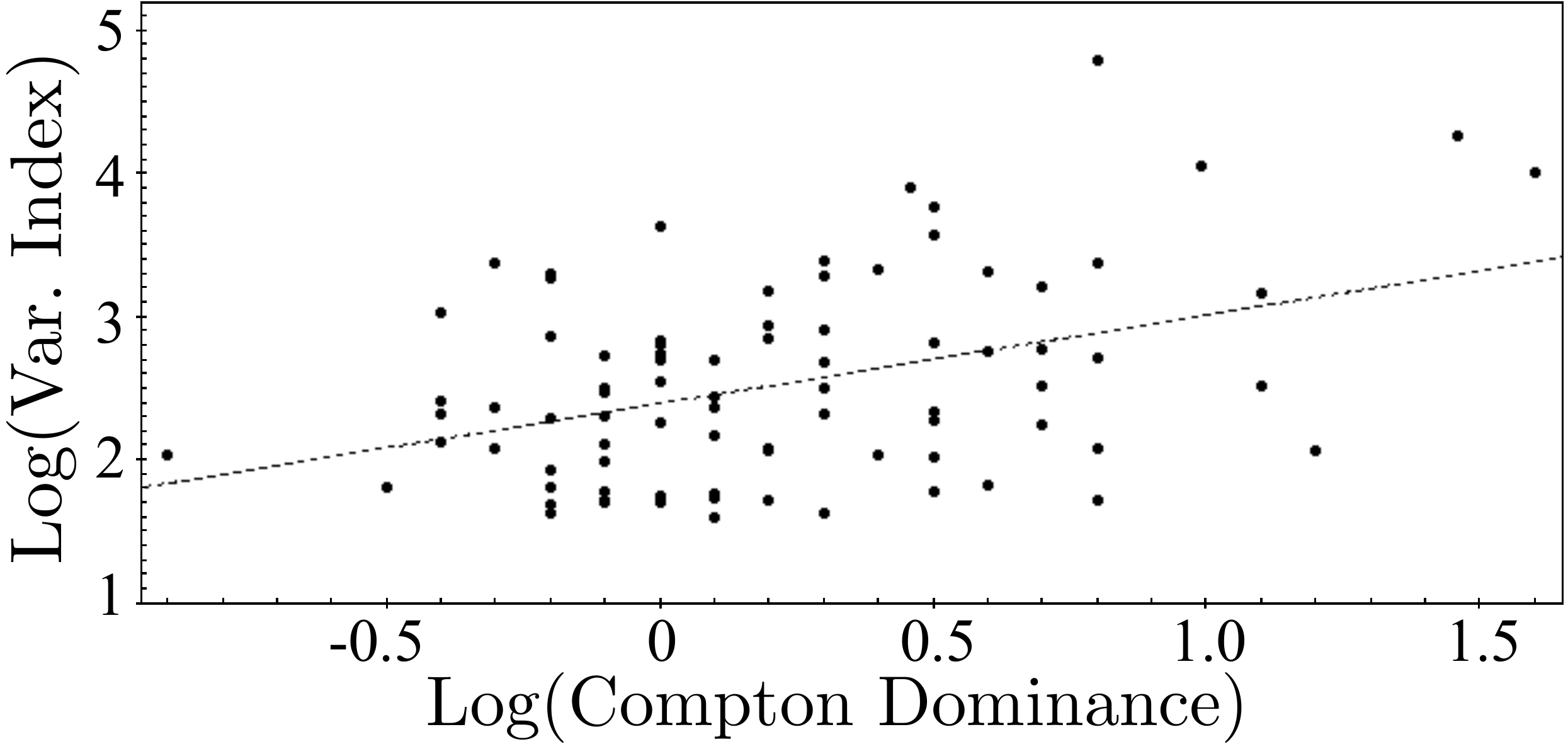}
     \caption{ \gr\ variability index vs. log(CD) for the Radio-Planck sample. The dashed line represents a linear fit log(variability Index) = m$\times$log(CD) + k, where the constants are m = 0.618 and k = 2.39.}
      \label{CDsubsample}
\end{figure}

According to \cite{3FGL} the variability index indicates if a $\gamma$-ray source is variable on a timescale of months, not addressing shorter or longer time variations. An index $>$\,72.4 indicates a $>$\,99\% confidence probability that the source is variable. At least $\approx$\,2/3 of the Radio-Planck sources (66 out of 104) have a variability index $>$\,72.4 therefore detected as variables on a timescale of months. To investigate if the variability index could be correlated to the CD, we plot in Fig. \ref{CDsubsample} the \gr\ $\rm \log(Var. index)$ taken from the 3FGL catalog versus our estimate of the parameter $\rm \log(CD)$. A linear fit of the scatter plot (Fig. \ref{CDsubsample}) has a Pearson correlation coefficient of 0.40, meaning the positive correlation between $\rm \log(Var. index)$ and $\rm \log(CD)$ is relatively weak. In fact, this only tells us that the variability index might not be the best parameter to rely on if we are willing to investigate the influence of $\gamma$-ray variability over the CD parameter. 

Looking at individual cases however provides a better picture of CD variations induced by fast variability. We consider first the three sources detected in $\gamma$-rays only during flaring episodes (Sec. \ref{flaring}). These objects move from log(CD)\,$<$\,$-$1.0 (during \gr\ quiet period) up to 0.79, 0.31, and 2.51, respectively, for BZQJ\,0010+1058, BZUJ\,0241-0815, and  BZQJ\,2136+0041, and all cases show variability on timescales at least lower than one month (given that they are detected within isolated month bins), i.e., shorter than the monthly time bin described by the variability index parameter. Further studies are necessary to investigate the dependencies of transient and fast flares with respect to a time-bin smaller than that of a month, which we used. 

\begin{figure}[h]
   \centering
    \includegraphics[width=1.0\linewidth]{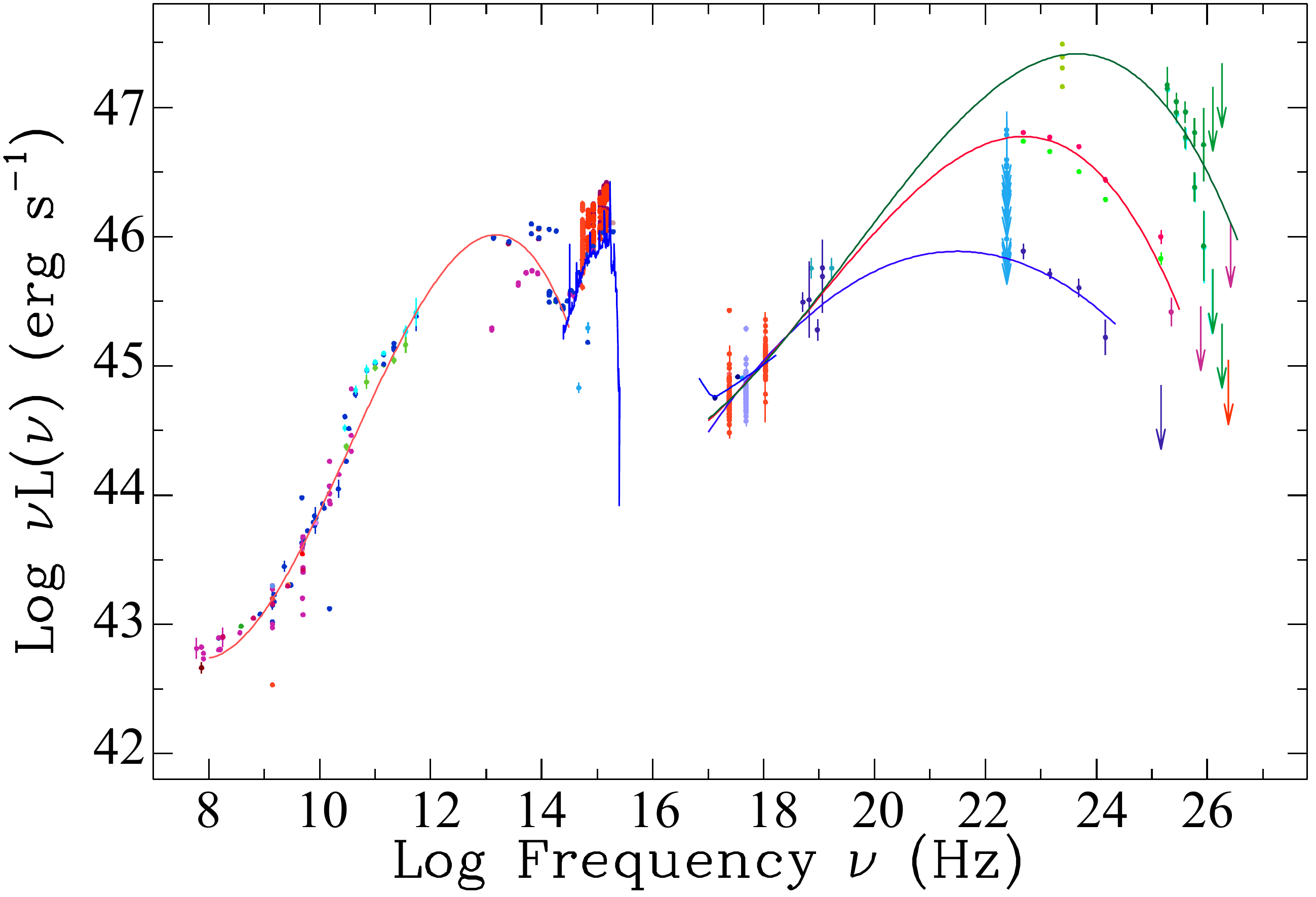}
     \caption{SED for BZQJ\,1224+2122, with blue bump template assuming z\,=\,0.434. This source has the largest \gr\ variability index, and we fit the IC component during various flaring states. In the high-energy band, the blue line refers to the 1FGL detection (integrating 2008-2009 data), the red line refers to the 3FGL detection (2008-2012 data), and the dark-green line represents a short and relatively bright flare reported by the MAGIC team \cite{MagicBZQ1224+2122} with reported variability within few hours.}
      \label{gammavar}
\end{figure}

Another example of strong \gr\ variability is BZQJ\,1224+2122 (4C\,+21.35), whose SED is shown in Fig.~\ref{gammavar}. We fit both Syn and IC components with a third order polynomial, listing fit parameters and their corresponding CD values in table \ref{var1}. We note that the high-energy peak flux changes by one order of magnitude in between 1FGL (dark blue) and 2FGL catalogs (red). This kind of long timescale variability (several months) is well represented by the variability index. When checking the light curve available online\footnote{Light curve for BZQJ\,1224+2122 (3FGLJ\,1224.9+2122): \url{http://fermi.gsfc.nasa.gov/ssc/data/access/lat/4yr_catalog/ap_lcs/lightcurve_3FGLJ1224.9p2122.png}} we see that this source had a relatively steady \gr\ emission during the first year of observations by Fermi-LAT (08/2008 to 08/2009, corresponding to the 1FGL SED; dark blue points), while it later underwent a strong activity when the photon flux varied by more than one order of magnitude in the 0.1-500\,GeV band. 

Since a more active state appears just after the integration time used for building the 1FGL catalog, we know that the flaring activity is now  smoothed over two and four years of integration time. These integration times are used for the 2FGL and 3FGL catalogs represented by red and light green points, respectively, in the high-energy SED. However, when integrating data only during the brightest state, it is possible to observe E\,$>$\,100\,GeV flux variability of $\approx$ one order of magnitude within hourly timescale, as for the flaring episode reported by \cite{MagicBZQ1224+2122}. This is a clear example of how the log(CD) parameter can vary widely, from $-$0.2 (during steady \gr\ emission) to +0.6 (when integrating over steady+flaring states), and reaching up to +1.3 (at the peak-flaring) as reported in Table \ref{var1}.

\begin{table}[h]
\centering
\caption{Inverse Compton peak parameters in various flaring states for 5BZQJ\,1224+2122. First to last line in table correspond to blue, red, and green curves, respectively, from Fig. \ref{gammavar}.}  
\label{var1}
\begin{tabular}{cccc}
Epoch & log($\nu^{peak}_{IC}$)   &  log($\nu$f$^{}_{\nu}$) &  log(CD)       
\\
\hline
1FGL & 21.5    &    $-$10.9   &   $-$0.2  \\
3FGL & 22.7    &    $-$10.1   &   +0.6    \\
Flare & 23.6    &    $-$9.40   &   +1.3    \\  
\end{tabular}
\end{table}

Overall, these examples suggest that \gr\ flaring states are likely to generate the largest CD values in the tail of the distribution (Fig. \ref{dominance}, top). During flare episodes, the sources could be moving from a nearly steady multicomponent SSC + EC regime, to a short-lived EC-dominated regime, which produce large amplitude variability (up to three orders of magnitude) owing to the extra beaming factor $\propto \delta^{1-\alpha}$ \cite{Dermer} that is present in EC scenario. We will investigate that in a forthcoming paper.

\subsection{Compton dominance versus ${\nu}^{Syn}_{peak}$ plane}

\begin{figure}[]
   \centering
    \includegraphics[width=1.0\linewidth]{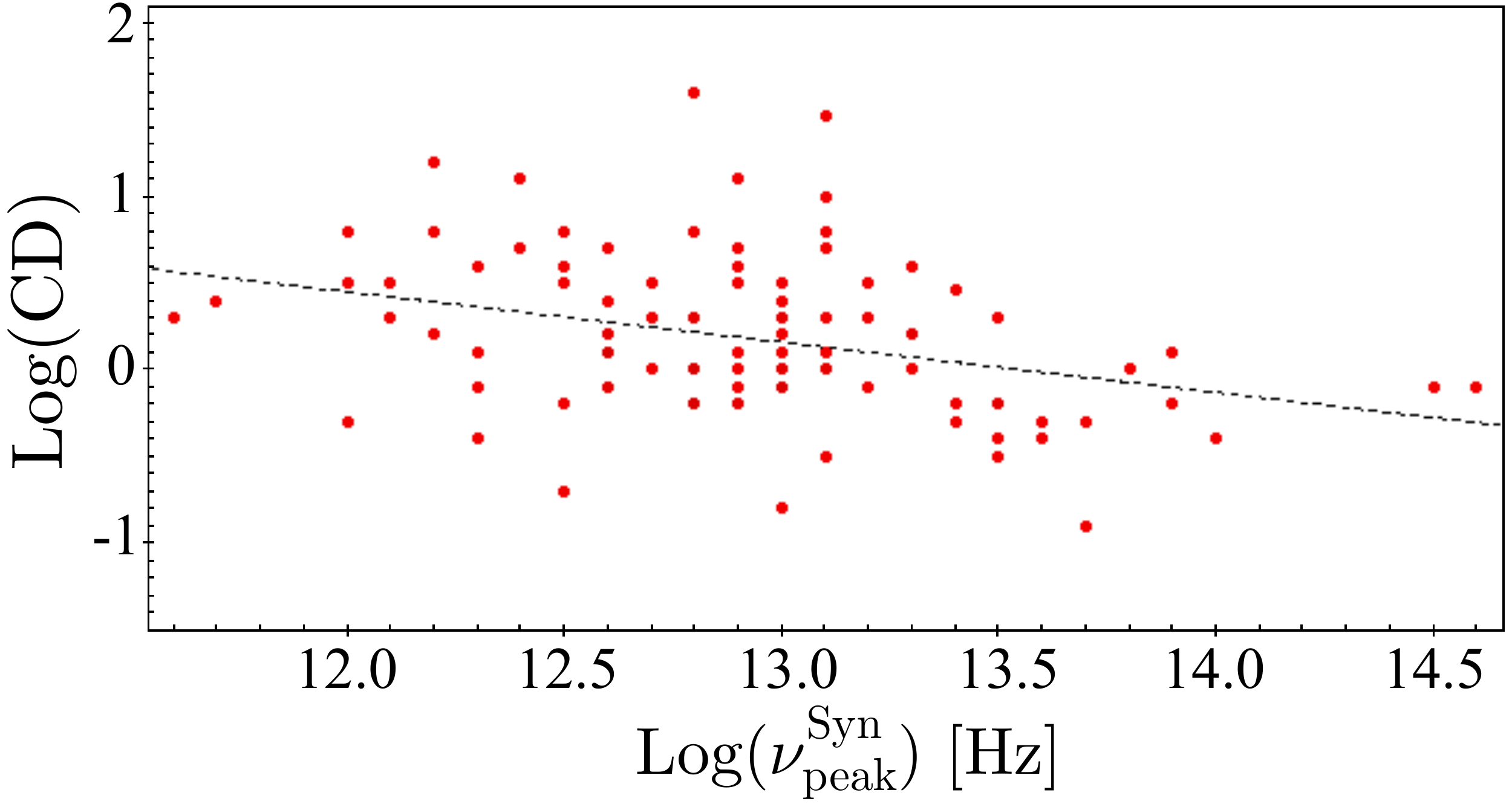}
     \caption{Log(CD) vs. log($\nu^{Syn}_{peak}$) plane for the Radio-Planck sample, probing almost three orders of magnitude in $\nu^{Syn}_{peak}$ space and showing very weak correlation with a large scatter. The black dashed line is a linear fit log(CD)\,=\,$-$0.29\,log($\nu^{Syn}_{peak}$)+3.92 .}
      \label{CDnu-peak}
\end{figure}

Here we report on the relation between log(CD) and log($\nu^{Syn}_{peak}$) that has been argued in literature \citep{Fossati-1998,CDsequency2017A&A,CDblazars} to show a relatively strong correlation. In fact, for our sample the Pearson's correlation coefficient {\it r} between log(CD) and log($\nu^{Syn}_{peak}$) is very weak, $r\sim$\,$-$0.336, a direct consequence of the large scatter associated with these parameters. As shown in figure \ref{CDnu-peak}, we cover almost three decades in $\nu_{peak}$ space, suggesting that the correlation between log(CD) and log($\nu_{peak}$) parameters should be considered with great care in order to evaluate its dependence on selection effects when building the study sample. 

It is also important to keep in mind that CD estimates reported for blazars with log($\nu_{peak}$)\,$>$\,15.0 are rather uncertain and subject to strong selection effects. It is extensively mentioned in literature \citep{3FGL,3LAC,1WHSP,2WHSP} that HSP blazars on average have a hard $\gamma$-ray spectral slope with $\langle \Gamma \rangle $ ranging from 1.8 to 2.0, and therefore the IC peak is most of the times out of reach for the Fermi-LAT, given its sensitivity window. For the brightest cases in which the IC peak was probed by VHE Cherenkov observatories, there is still the uncertainty introduced by absorption of VHE photons due to pair creation when scattering EBL photons. Indeed, only a few HSP sources have their IC peak probed by VHE observatories with observations triggered by X-ray and $\gamma$-ray flaring states. This alone introduces a strong bias, given that no VHE blind sky survey is available.

\section{Conclusions}

The Radio-Planck sample includes 104 bright radio-selected sources (f $> $1 Jy at 37\,GHz), 102 of which are optically identified blazars in the 5BZcat and were expected to be detected by the Fermi-LAT. The two remaining objects are radio galaxies, namely 3C111 and M87, both detected in $\gamma$-rays. The noninclusion of a fair fraction of 5BZcat sources in published Fermi-LAT catalogs motivated our search for new \gr\ detections using 7.5 years of data as available at the time of writing. The main results of our work can be summarized as follows. Out of 104 sources, 83 have counterparts from FGL catalogs (all TS$>$25); 6 are new detections with TS\,$>$\,20; 3  are new detections with TS in between 10 to 20; 3 are new associations with 3FGL sources  (from improved positioning with high-energy TS maps); 1  is a new detection from solving \gr\ source-confusion; and 3 are transients that were detected during short flaring episodes. 

Five sources remain undetected in the \gr\, band, all of which are optically identified blazars included in the 5BZcat. Two objects have relatively high redshift (5BZQJ\,2139+1423 at z\,=\,2.427 and 5BZQJ\,2007+4029 at z\,=\,1.736). The remaining three sources are 5BZQJ\,0927+3902 (z\,=\,0.695), 5BZQJ\,2022+6136 (z\,=\,0.228), 5BZQJ\,1927+7358 (z\,=\,0.302).


We conclude that most of sources currently called \gr\ quiet blazars are actually associated with relevant \gr\ signatures, becoming evident by means of a dedicated case-by-case study of 7.5 yr of Fermi-LAT observations. At most, \gr\ quiet blazars might be a very small fraction of the LSP population, suggesting there is no urgent need to introduce a new blazar class. From the five nondetections reported, two are high redshift sources, where absorption may hinder a \gr\ signature. Another case (5BZQJ\,0927+3902) is associated with a bright 3FGL source that dominates the region, such that we could not probe for source confusion. Finally, 5BZQJ\,2022+6136 and 5BZQJ\,1927+7358 only showed hints of flaring activity and are probably under Fermi-LAT sensitivity.

All new detections reported in this work contribute to solve a small fraction of the extragalactic \gr\ background into point-like sources. We note that the presence of transient sources, which are only detectable during short flaring episodes, could represent a non-negligible fraction of the MeV to GeV background. This would be a possible approach to consider in future studies. We discuss examples of how to extract refined and relevant \gr\ information by considering a multifrequency approach when searching for new sources, showing that Fermi-LAT database is a large resource still to be explored in detail.

We study the CD distribution, showing that a single Gaussian function fails to describe the cases with large log(CD) values at the tail of the histogram. There is indeed a number of high log(CD) sources that are in excess with respect to a single Gaussian fitting. We evaluate the impact of fast \gr\ variability on the CD parameter, considering 5BZQJ\,1224+2122 as an example. We point out three cases in which large CD values are observed during fast flaring states (BZQJ\,0010+1058, BZUJ\,0241-0815, and  BZQJ\,2136+0041), such that CD values can be one to two orders of magnitude larger compared to those obtained during the steady and relatively faint \gr\ emission. As follows, the absent correlation between log(CD) versus log(Var. Index) in the scatter plot from Fig.~\ref{CDsubsample} shows that the $\gamma$-ray variability index may not be the best tool to evaluate the relation connecting $\gamma$-ray flaring states and large log(CD) sources. Finally, we also evaluate the putative correlation between log(CD) and log($\nu^{Syn}_{peak}$) parameters, finding relatively weak evidence for that. The similarity between Syn and IC $\nu$f$_{\nu}$ peak distributions and the tight peak ratio log($\nu^{IC}_{peak}$/$\nu^{Syn}_{peak}$) distribution points to a dominant mechanism (either SSC or EC) to account for the IC component in bright LSP blazars, otherwise we would have found a large spread in the parameter space we probed. An extensive evaluation testing SSC and EC scenarios is explored with great detail in a parallel work \citep{paperII}, which is based on the Radio-Planck sample.

Also, we showed a few examples for which the power-law fitting parameters estimated for faint $\gamma$-ray blazars, which were detected with TS between 10 to 25 under 3FGL setup integrating over 4.0 years of Pass7 data, are later confirmed when integrating over a larger exposure time of 7.5 years. This vindicates the importance and usefulness of reporting faint \gr\ signatures in association with blazar counterparts.  

\begin{acknowledgements}
     
During this work, BA was supported by the Brazilian Scientific Program Ci\^{e}ncias sem Fronteiras - Cnpq, and later by S\~ao Paulo Research Foundation (FAPESP) with grant n. 2017/00517-4. We would like to thank Prof. Paolo Giommi for his comments along the preparation of this work, Prof. Marcelo M. Guzzo and Prof. Orlando L. G. Peres for the full support which allowed the author partnership with FAPESP. We thanks IcraNet and Prof. Carlo Bianco for the cooperation granting access to Joshua Computer Cluster (Rome-Italy) for Fermi-LAT data reduction. We thank the CCJDR Data Center at IFGW Unicamp (Campinas-Brazil) where we also performed Fermi-LAT data reduction at their Feynman Cluster. We thank SSDC, Space Science Data Center from Agenzia Spaziale Italiana; University La Sapienza of Rome, Department of Physics; And State University of Campinas - Unicamp, IFGW Department of Physics for hosting the author. We make use of archival data and bibliographic information obtained from the NASA-IPAC Extragalactic Database (NED), data, and software facilities from the SSDC.

\end{acknowledgements}

\bibliographystyle{aa}
\bibliography{Fermipaper}

\begin{longtable}{llrcc|cccc}
\caption{Here we lits all 104 sources used for our studies. Column 5BZcat shows the blazar name according to \cite{5BZcat}, where BZQ stands for Flat Spectrum Radio Quasars, BZB for BL Lacs and BZU for still undefined-class blazars. Following we list coordinates R.A. and Dec. (J2000) for each source. Columns NVSS shows the radio counterpart according to the NVSS catalog \citep{NVSS}. Column z corresponds to the redshift as reported in the 5BZcat \cite{5BZcat}, and from the NASA/IPAC Extragalactic Database (NED). Columns $ \rm log(\nu^{Syn}_{peak})$ and log($\nu^{Syn}_{peak}$) are the fitting parameters referring to the peak frequency from Syn and IC components measured in Hz. Columns log($\nu f^{Syn}_{\nu}$) and log($\nu f^{IC}_{\nu}$) correspond to the Syn and IC peak-power measured in erg/cm$^2$/s. All fitting parameters are given as a measure of the mean SED, considering all available data.}\\
\hline\hline
5BZcat\,J          & R.A.        &  Dec.     &   NVSS\,J  & z   & log($\nu$) & log($\nu$f$_{\nu}$) & log($\nu_{IC}$) &  log($\nu$f$_{\nu-IC}$)  \\
\hline
\endfirsthead
\caption{continued.}\\
\hline\hline
5BZcat\,J          & R.A.        &  Dec.     &   NVSS\,J  & z   & log($\nu$) & log($\nu$f$_{\nu}$) & log($\nu_{IC}$) &  log($\nu$f$_{\nu-IC}$)  \\
\hline
\endhead
\hline
\endfoot
   5BZBJ0050-0929   &      12.67167 &    -9.48500 &  005041-092906    &    0      &  14.6   &-11.0 & 22.7   &  -11.1       \\ 
   5BZQJ1512-0905   &     228.21056 &    -9.09995 &  151250-090600    &    0.360  &  13.1   &-10.9 & 22.2   &  -9.91       \\  
   5BZQJ2229-0832   &     337.41705 &    -8.54847 &  222940-083254    &    1.560  &  13.1   &-11.2 & 21.8   &  -10.5       \\  
   5BZQJ0607-0834   &      91.99875 &    -8.58056 &  060759-083450    &    0.870  &  12.1   &-11.5 & 21.4   &  -11.0       \\   
   5BZUJ0241-0815   &      40.27000 &    -8.25576 &  024104-081521    &    0.005  &  13.5   &-10.1 & 20.9(?)&  -10.6       \\      
   5BZQJ0808-0751   &     122.06473 &    -7.85275 &  080815-075109    &    1.837  &  13.0   &-11.1 & 22.9   &  -10.7       \\       
   5BZQJ0006-0623   &       1.55791 &    -6.39333 &  000613-062335    &    0.347  &  13.0   &-11.1 & 20.3   &  -11.9       \\       
   5BZQJ1256-0547   &     194.04652 &    -5.78931 &  125611-054720    &    0.536  &  12.8   &-10.0 & 22.7   &  -10.0       \\      
   5BZQJ2225-0457   &     336.44693 &    -4.95039 &  222547-045701    &    1.404  &  13.0   &-10.8 & 21.7   &  -10.9       \\      
   5BZQJ2218-0335   &     334.71683 &    -3.59358 &  221852-033537    &    0.901  &  12.3   &-11.3 & 21.0   &  -11.7       \\      
   5BZQJ1743-0350   &     265.99527 &    -3.83461 &  174358-035004    &    1.057  &  12.6   &-11.3 & 21.0   &  -11.2       \\        
   5BZBJ2134-0153   &     323.54294 &    -1.88812 &  213410-015317    &    1.283  &  12.8   &-11.3 & 21.8   &  -11.5       \\        
   5BZQJ0501-0159   &      75.30338 &    -1.98729 &  050112-015912    &    2.291  &  13.0   &-11.4 & 21.4   &  -11.1       \\      
   5BZQJ0339-0146   &      54.87891 &    -1.77667 &  033930-014635    &    0.805  &  12.6   &-11.3 & 22.0   &  -11.1       \\         
   5BZQJ0423-0120   &      65.81583 &    -1.34253 &  042315-012032    &    0.916  &  13.0   &-10.6 & 22.1   &  -10.7       \\         
   5BZUJ0725-0054   &     111.46125 &    -0.91556 &  072550-005458    &    0.128  &  13.5   &-11.0 & 20.5   &  -11.2       \\         
   5BZQJ0125-0005   &      21.37017 &    -0.09889 &  012528-000556    &    1.077  &  12.8   &-11.6 & 20.3   &  -11.6       \\         
   5BZQJ2136+0041   &     324.16080 &     0.69839 &  213638+004154    &    1.941  &  11.7   &-11.6 & 21.2   &  -11.2       \\        
   5BZQJ0108+0135   &      17.16154 &     1.58342 &  010838+013458    &    2.099  &  12.9   &-11.2 & 22.4   &  -10.6       \\       
   5BZQJ0739+0137   &     114.82513 &     1.61794 &  073918+013704    &    0.189  &  13.9   &-10.9 & 21.6   &  -10.8       \\       
   5BZUJ1058+0133   &     164.62338 &     1.56633 &  105829+013358    &    0.890  &  13.1   &-10.9 & 22.3   &  -10.8       \\       
   5BZQJ1229+0203   &     187.27792 &     2.05222 &  122906+020305    &    0.158  &  13.4   &-10.0 & 20.8   &  -9.54       \\         
   5BZQJ1549+0237   &     237.37267 &     2.61700 &  154929+023700    &    0.414  &  12.9   &-11.2 & 22.0   &  -11.1       \\           
   5BZBJ0825+0309   &     126.45958 &     3.15667 &  082550+030924    &    0.506  &  13.1   &-11.1 & 21.5(?)&  -11.6       \\           
   5BZQJ1222+0413   &     185.59375 &     4.22083 &  122222+041317    &    0.966  &  12.7   &-11.2 & 20.8   &  -10.7       \\           
   5BZUJ0433+0521   &      68.29620 &     5.35433 &  043311+052115    &    0.033  &  13.8   &-10.2 & 19.8   &  -10.2       \\         
   5BZQJ2123+0535   &     320.93549 &     5.58947 &  212344+053522    &    1.941  &  12.6   &-11.7 & 21.6   &  -11.8       \\            
   5BZQJ1038+0512   &     159.69492 &     5.20808 &  103846+051229    &    0.473  &  12.0   &-11.8 & 20.8   &  -12.1       \\          
   5BZQJ1550+0527   &     237.64696 &     5.45290 &  155035+052710    &    1.422  &  13.0   &-11.4 & 21.8   &  -11.4       \\
   5BZQJ2148+0657   &     327.02252 &     6.96055 &  214805+065739    &    0.999  &  12.5   &-10.8 & 20.5   &  -11.0       \\            
   5BZUJ1751+0939   &     267.88675 &     9.65019 &  175132+093901    &    0.322  &  13.1   &-10.8 & 21.9   &  -10.8       \\           
   5BZBJ0757+0956   &     119.27766 &     9.94303 &  075706+095634    &    0.266  &  13.7   &-10.8 & 20.8   &  -11.1       \\            
   5BZQJ0309+1029   &      47.26500 &    10.48778 &  030903+102916    &    0.863  &  12.8   &-11.2 & 21.7   &  -11.2       \\              
   5BZQJ1608+1029   &     242.19249 &    10.48550 &  160846+102908    &    1.226  &  12.8   &-11.3 & 21.6   &  -11.0       \\               
   5BZQJ1504+1029   &     226.10408 &    10.49417 &  150425+102938    &    1.839  &  12.8   &-11.6 & 22.9   &  -10.0       \\            
   5BZQJ0010+1058   &       2.62917 &    10.97472 &  001030+105827    &    0.089  &  14.5   &-10.7 & 20.5   &  -10.8       \\                
   5BZBJ0449+1121   &      72.28196 &    11.35778 &  044907+112128    &    2.153  &  12.9   &-11.5 & 21.8   &  -10.8       \\         
   5BZQJ2232+1143   &     338.15167 &    11.73055 &  223236+114350    &    1.037  &  12.4   &-11.2 & 21.3   &  -10.5       \\         
   5BZQJ0750+1231   &     117.71687 &    12.51801 &  075052+123104    &    0.889  &  12.6   &-11.1 & 21.2   &  -11.2       \\           
   5BZQJ0530+1331   &      82.73508 &    13.53200 &  053056+133155    &    2.070  &  12.2   &-11.5 & 21.4   &  -10.7       \\         
   5BZUJ1415+1320   &     213.99500 &    13.34000 &  141558+132024    &    0.247  &  12.8   &-11.0 & 20.5   &  -11.2       \\         
   5BZQJ2139+1423   &     324.75546 &    14.39333 &  213901+142336    &    2.427  &  12.2   &-11.7 & -      &  -           \\            
   5BZUJ0204+1514   &      31.21004 &    15.23639 &  020450+151411    &    0.833  &  12.6   &-11.6 & 21.0   &  -11.2       \\           
   5BZQJ2253+1608   &     343.49042 &    16.14805 &  225357+160853    &    0.859  &  13.1   &-10.0 & 22.2   &  -9.20       \\                
   5BZBJ0238+1636   &      39.66167 &    16.61639 &  023838+163658    &    0.940  &  13.0   &-10.9 & 22.5   &  -10.4       \\          
   5BZQJ2203+1725   &     330.86203 &    17.43006 &  220326+172548    &    1.076  &  13.3   &-10.9 & 22.9   &  -10.9       \\           
   5BZBJ0738+1742   &     114.53083 &    17.70528 &  073807+174219    &    0.424  &  13.5   &-10.6 & 23.8   &  -11.0       \\            
   5BZQJ0510+1800   &      77.50988 &    18.01155 &  051002+180041    &    0.416  &  13.3   &-11.3 & 21.3   &  -11.1       \\             
   5BZBJ0854+2006   &     133.70332 &    20.10833 &  085448+200630    &    0.306  &  13.6   &-10.4 & 21.8   &  -10.8       \\           
   5BZQJ1224+2122   &     186.22713 &    21.37972 &  122454+212247    &    0.434  &  13.1   &-10.7 & 23.0   &  -10.0       \\          
   5BZQJ0152+2207   &      28.07525 &    22.11881 &  015218+220707    &    1.320  &  12.9   &-11.5 & 21.8   &  -11.5       \\         
   5BZQJ1327+2210   &     201.75359 &    22.18060 &  132700+221050    &    1.398  &  12.6   &-11.7 & 21.7   &  -11.0       \\          
   5BZQJ0830+2410   &     127.71700 &    24.18333 &  083052+241058    &    0.939  &  12.6   &-11.1 & 21.8   &  -11.0       \\         
   5BZQJ1043+2408   &     160.78749 &    24.14306 &  104309+240835    &    0.560  &  12.9   &-11.5 & 21.7   &  -11.6       \\          
   5BZQJ0956+2515   &     149.20784 &    25.25446 &  095649+251515    &    0.712  &  12.7   &-11.4 & 21.9   &  -11.4       \\          
   5BZQJ2236+2826   &     339.09363 &    28.48261 &  223622+282858    &    0.790  &  13.0   &-11.2 & 22.5   &  -10.9       \\           
   5BZQJ0237+2848   &      39.46838 &    28.80250 &  023752+284809    &    1.206  &  12.9   &-11.2 & 22.0   &  -10.7       \\              
   5BZQJ1159+2914   &     179.88263 &    29.24556 &  115931+291444    &    0.729  &  13.5   &-11.1 & 22.6   &  -10.8       \\          
   5BZQJ2203+3145   &     330.81210 &    31.76056 &  220314+314538    &    0.295  &  13.4   &-10.9 & 20.9   &  -11.1       \\         
   5BZQJ0336+3218   &      54.12542 &    32.30806 &  033630+321829    &    1.259  &  12.8   &-11.3 & 20.2   &  -10.5       \\         
   5BZUJ1310+3220   &     197.61940 &    32.34550 &  131028+322044    &    0.997  &  13.1   &-10.8 & 22.2   &  -10.8       \\              
   5BZQJ1613+3412   &     243.42084 &    34.21333 &  161341+341247    &    1.397  &  12.3   &-11.5 & 21.7   &  -11.6       \\                  
   5BZQJ1635+3808   &     248.81454 &    38.13458 &  163515+380804    &    1.814  &  12.5   &-11.1 & 21.7   &  -10.3       \\               
   5BZQJ1130+3815   &     172.72198 &    38.25514 &  113053+381519    &    1.733  &  12.3   &-11.7 & 22.6   &  -11.6       \\           
   5BZQJ0927+3902   &     141.76254 &    39.03914 &  092703+390220    &    0.695  &  12.1   &-11.2 & -      &  -           \\           
   5BZQJ0555+3948   &      88.87837 &    39.81366 &  055530+394848    &    2.365  &  12.0   &-11.7 & 20.8   &  -10.9       \\           
   5BZBJ1653+3945   &     253.46750 &    39.76000 &  165352+394536    &    0.033  &  17.8   &-10.2 & 25.2   &  -10.6       \\            
   5BZQJ1640+3946   &     250.12347 &    39.77898 &  164029+394646    &    1.660  &  12.9   &-11.9 & 22.7   &  -10.8       \\          
   5BZQJ1642+3948   &     250.74506 &    39.81028 &  164258+394837    &    0.593  &  13.0   &-10.6 & 21.7   &  -10.7       \\         
   5BZQJ2007+4029   &     301.93726 &    40.49683 &  200744+402948    &    1.736  &  12.2   &-11.6 & -      &  -           \\             
   5BZQJ0948+4039   &     147.23059 &    40.66239 &  094855+403944    &    1.249  &  12.3   &-11.8 & 20.8   &  -11.2       \\             
   5BZUJ0319+4130   &  *   49.95042 &    41.51167 &  031948+413042    &    0.018  &  13.0(?)&-10.4 & 23.3   &  -10.4       \\           
   5BZBJ2202+4216   &     330.68042 &    42.27750 &  220243+421640    &    0.069  &  13.6   &-10.0 & 21.3   &  -10.3       \\            
   5BZUJ0909+4253   &     137.38959 &    42.89613 &  090933+425347    &    0.670  &  12.9   &-11.5 & 20.9   &  -11.7       \\           
   5BZQJ0646+4451   &     101.63346 &    44.85461 &  064632+445116    &    3.396  &  11.6   &-11.8 & 21.8(?)&  -11.5       \\         
   5BZQJ0920+4441   &     140.24359 &    44.69833 &  092058+444153    &    2.190  &  12.5   &-11.2 & 22.3   &  -10.6       \\            
   5BZQJ2354+4553   &     358.59033 &    45.88445 &  235421+455304    &    1.992  &  12.2   &-11.9 & 21.4   &  -11.7       \\         
   5BZQJ0136+4751   &      24.24412 &    47.85809 &  013658+475129    &    0.859  &  13.3   &-10.9 & 22.2   &  -10.7       \\           
   5BZUJ1829+4844   &     277.38251 &    48.74628 &  182931+484446    &    0.695  &  13.0   &-11.3 & 20.7   &  -11.1       \\            
   5BZQJ1153+4931   &     178.35196 &    49.51911 &  115324+493109    &    0.334  &  12.9   &-10.9 & 21.2   &  -11.1       \\                      
   5BZQJ0808+4950   &     122.16529 &    49.84347 &  080839+495036    &    1.432  &  12.0   &-12.2 & 20.6   &  -11.7       \\           
   5BZQJ0359+5057   &      59.87396 &    50.96394 &  035929+505750    &    1.512  &  12.1   &-10.7 & 21.3   &  -10.4       \\               
   5BZQJ2038+5119   &     309.65433 &    51.32018 &  203837+511913    &    1.686  &  12.5   &-11.5 & 21.4   &  -11.0       \\          
   5BZQJ1955+5131   &     298.92810 &    51.53015 &  195542+513149    &    1.214  &  12.7   &-11.7 & 20.7   &  -11.4       \\          
   5BZQJ1740+5211   &     265.15408 &    52.19528 &  174036+521143    &    1.381  &  13.2   &-11.3 & 21.3   &  -10.8       \\              
   5BZBJ1419+5423   &     214.94417 &    54.38722 &  141946+542315    &    0.153  &  13.7   &-10.6 & 21.9   &  -11.5       \\            
   5BZBJ1824+5651   &     276.02948 &    56.85042 &  182407+565101    &    0.663  &  13.2   &-11.3 & 22.1   &  -11.0       \\               
   5BZUJ0102+5824   &      15.69067 &    58.40310 &  010245+582411    &   ?0.644  &  12.6   &-11.1 & 21.8   &  -10.9       \\              
   5BZQJ2022+6136   &     305.52783 &    61.61633 &  202206+613658    &    0.228  &  12.9   &-11.2 & -      &  -           \\            
   5BZBJ0958+6533   &     149.69666 &    65.56499 &  095847+653354    &    0.367  &  13.4   &-11.0 & 21.2   &  -11.3       \\             
   5BZQJ1849+6705   &     282.31833 &    67.09403 &  184915+670540    &    0.657  &  13.1   &-10.9 & 22.5   &  -10.6       \\           
   5BZQJ0228+6721   &      37.20854 &    67.35084 &  022850+672101    &    0.523  &  12.8   &-11.2 & 21.1   &  -11.4       \\             
   5BZQJ1642+6856   &     250.53271 &    68.94437 &  164207+685638    &    0.751  &  12.5   &-11.6 & 20.1(?)&  -12.3       \\         
   5BZBJ1806+6949   &     271.71124 &    69.82445 &  180650+694928    &    0.046  &  14.0   &-10.6 & 21.7   &  -11.0       \\            
   5BZQJ0841+7053   &     130.35153 &    70.89506 &  084124+705341    &    2.218  &  12.4   &-11.3 & 20.1   &  -10.2       \\                  
   5BZBJ0721+7120   &     110.47208 &    71.34333 &  072153+712036    &    0.0    &  13.9   &-10.3 & 23.3   &  -10.5       \\         
   5BZQJ0217+7349   &      34.37833 &    73.82555 &  021730+734932    &    2.367  &  12.2   &-11.6 & 20.5   &  -10.4       \\           
   5BZQJ1927+7358   &     291.95209 &    73.96711 &  192748+735802    &    0.302  &  13.1   &-10.8 & -      &  -           \\              
   5BZBJ2005+7752   &     301.37961 &    77.87861 &  200531+775243    &    0.342  &  13.2   &-11.2 & 21.5   &  -11.3       \\                
   5BZBJ1800+7828   &     270.19034 &    78.46778 &  180045+782805    &    0.680  &  13.5   &-10.7 & 21.9   &  -10.9       \\           
   5BZQJ1153+8058   &     178.30208 &    80.97476 &  115312+805829    &    1.250  &  12.6   &-12.0 & 21.1   &  -11.9       \\           
   5BZQJ2356+8152   &     359.09497 &    81.88118 &  235622+815252    &    1.344  &  12.8   &-11.8 & 21.1   &  -11.5       \\         
         M87        &     187.70592 &    12.39111 &  123049+122321    &    0.0042 &  13.0   &-10.5 & 18.5(?)&  -10.4       \\              
         3C111      &      64.58867 &    38.02661 &  041820+380148    &    0.0485 &  13.3   &-10.6 & 20.1   &  -10.0       \\

\label{tableRadioPlanck}
\end{longtable}


\end{document}